\documentstyle[aps,prd,epsf]{revtex}

\newcommand{\Aslash}{\not{\!\! A}}
\newcommand{\pslash}{\not{\! p}}

\newcommand{\bfGamma}{{\bf \Gamma}}
\newcommand{\bftGamma}{\tilde{\bf \Gamma}}

\newcommand{\bfchi}{\bbox{\chi}}
\newcommand{\bflam}{\bbox{\lambda}}
\newcommand{\bfeta}{\bbox{\eta}}
\newcommand{\bfsig}{\bbox{\sigma}}

\newcommand{\bftchi}{\tilde{\bbox{\chi}}}
\newcommand{\bftlam}{\tilde{\bbox{\lambda}}}
\newcommand{\bfteta}{\tilde{\bbox{\eta}}}
\newcommand{\bftsig}{\tilde{\bbox{\sigma}}}

\newcommand{\be}{\begin{equation}}
\newcommand{\ee}{\end{equation}}
\newcommand{\ba}{\begin{eqnarray}}
\newcommand{\ea}{\end{eqnarray}}

\preprint{UCTP-105-00}

\begin{document}
\draft

\title{Diquarks in cold dense QCD with two flavors}

\author{V.A.~Miransky}
\address{Bogolyubov Institute for Theoretical Physics,
         252143, Kiev, Ukraine\\ 
     and Department of Physics, Nagoya University,
         Nagoya 464-8602, Japan}
\author{I.A.~Shovkovy\thanks{On leave of absence from 
                     Bogolyubov Institute for Theoretical 
                     Physics, 252143, Kiev, Ukraine.}
                     and
        L.C.R.~Wijewardhana}

\address{Physics Department, University of Cincinnati, 
         Cincinnati, Ohio 45221-0011}

\date{\today}
\maketitle

\begin{abstract}
We derive and analyze the Bethe-Salpeter equations for spin zero
diquarks in the color superconducting phase of cold dense QCD with two
massless flavors. The spectrum of diquarks contains an infinite number
of massive excitations and five (nearly) massless pseudoscalars. The
former are singlets while the latter include a doublet, an antidoublet
and a singlet with respect to the unbroken $SU(2)_{c}$. Because of an
approximate parity doubling at large chemical potential, all massive
states come in pairs. The decay constants, as well as the velocities
of the (nearly) massless pseudoscalars are derived. The different role
of the Meissner effect for tightly bound states and quasiclassical
bound states is revealed. 
\end{abstract}

\pacs{11.10.St, 11.15.Ex, 12.38.Aw, 21.65.+f}



\section{Introduction}

With continuing advances in modern nuclear and high energy
experiments, it has become feasible to produce deconfined quark matter
in the laboratory. Not surprisingly, this has stimulated many
theoretical studies of quark matter at high densities and/or
temperatures (for recent reviews see, for example,
Refs.~\cite{Wilczek,short-reviews,Inst-Rev,Smilga}).

Of special interest is the cold quark matter at high densities, i.e.,
at densities which are at least a few times larger than that of a
nucleon $n_0\simeq 0.17 \mbox{~fm}^{-3}$. It was known for a long time
that such matter could be a color superconductor \cite{BarFra,Bail}.
Nevertheless, until recently our understanding of the color
superconducting state had remained very poor. The new developments
started with the ground breaking estimates of the color
superconducting order parameter in Refs.~\cite{W1,S1}. Within a
phenomenological (instanton liquid) model, it was shown there that the
order parameter could be as large as 100 MeV. These estimates revived
the hope of producing and detecting the color superconducting phase
either in experiments or in natural systems such as neutron (quark)
stars. Being motivated by the potential possibility of observing the
color superconducting phase at moderate densities, the subject bursted
with numerous studies and new discoveries. 

At first, it was not clear at all that the microscopic theory,
quantum chromodynamics (QCD), would lead to the same (or, at least,
the same order of magnitude) estimates for the superconducting gap as
the phenomenological models. It was suggested in Refs.~ \cite{PR1,Son}
that the screening effects of gluons should play a crucial role in the
analysis. In particular, while the electric gluon modes are subject to
the Debye screening at already the scales of order $l_{D}\sim
1/g_{s}\mu$, where $\mu$ is the chemical potential and $g_{s}$ is the
running coupling related to the scale $\mu$, the magnetic modes are
subject only to the Landau damping, which does not completely
eliminate the long range interaction \cite{Son}. The subsequent
studies of the gap equation in QCD confirmed that proper treatment of
the gluon screening effects are crucial in deriving the estimates for
the superconducting gap \cite{us,SW2,PR2,H1,Brown1,us2,Sch-patt}. 

Also, it was revealed that the ground state of the quark matter with
three light flavors is given by the so called Color Flavor Locked
(CFL) phase \cite{CFL}. It is remarkable that the chiral symmetry in
such a phase is spontaneously broken and most of the quantum numbers
of physical states coincide with those in the hadronic phase. It was 
tempting, therefore, to suggest that there might exist some kind of
continuity between the two phases \cite{Contin}. Another interesting
feature of the three flavor QCD was pointed out in
Ref.~\cite{gapless}, where the possibility of gapless color
superconductivity (a metastable phase) was proposed. In addition, many
interesting patterns of symmetry breaking were revealed in models with
the number of flavors larger than three \cite{Sch-patt}, as well as in
two-color QCD with the quarks in the fundamental representation and in
any-color QCD with quarks in the adjoint representation
\cite{2-color}. The anomaly matching conditions were  analyzed in
Ref.~\cite{Sannino}.

The low energy dynamics of the color superconducting phase could be
efficiently studied by using effective actions whose general structure
is fixed by symmetries \cite{SonSt,CasGat,Rho,HZB}. The finite set of
parameters in such theories could be either taken from an experiment
(when available), or sometimes derived from QCD (for example, in the
limit of the asymptotically large chemical potential). Because of the
nature of such an approach, at best it could probe the properties of
the pseudo-NG bosons, but not the detailed  spectrum of the diquark
bound states (mesons). It was argued in Ref.~\cite{us3}, however,
that, because of long-range interactions mediated by the gluons of the
magnetic type \cite{PR1,Son}, the presence of an infinite tower of
massive diquark states could be the key signature of the color
superconducting phase of dense quark  matter. 

In this paper, we consider the problem of spin zero bound states in
the two flavor color superconductor using the Bethe-Salpeter (BS)
equations. (A brief outline of our results was given in
Ref.~\cite{us-short}). We find that the spectrum contains five
(nearly) massless states and an infinite tower of massive singlets
with respect to the unbroken $SU(2)_{c}$ subgroup. Furthermore, the
following mass formula is derived for the singlets:
\begin{equation}
M^{2}_{n} \simeq 4 |\Delta^{-}_{0} |^{2} 
\left(1-\frac{\alpha_{s}^{2}\kappa}{(2n+1)^{4}}\right),
\quad n=1,2,\ldots,
\label{mass-singlet}
\end{equation}
where $\kappa$ is a constant of order 1 (we find that
$\kappa\simeq 0.27$),
$|\Delta^{-}_{0} |$ is the
dynamical Majorana mass of quarks in the color superconducting phase,
and $\alpha_{s}=g_{s}^2/{4\pi}$.

The Meissner effect plays a crucial role in obtaining this result. In
particular, the important point is that while the Meissner effect is
essentially irrelevant for tightly bound states, it is crucial for the
dynamics of quasiclassical bound states (whose binding energy is
small).

At large chemical potential, we also notice an approximate degeneracy
between scalar and pseudoscalar channels. As a result  of this parity
doubling, the massive diquark states come in pairs. In addition, there
also exist five massless scalars and five (nearly) massless
pseudoscalars [a doublet, an antidoublet and a singlet  under
$SU(2)_{c}$]. While the scalars are removed from the spectrum  of
physical particles by the Higgs mechanism, the pseudoscalars  remain
in the spectrum, and they are the relevant degrees of freedom  of the
infrared dynamics. At  high density, the massive and  (nearly)
massless states are narrow resonances.

This paper is organized as follows. In Sec.~\ref{model}, we describe
the model and introduce the notation. Then, further developing our
notation in Sec.~\ref{SD-equation}, we briefly review the approach of
the Schwinger-Dyson equation in the color superconducting phase of
$N_{f}=2$ QCD. In Sec.~\ref{Ward-id}, we derive the Ward identities
for the quark-gluon vertex functions, corresponding to the broken
generators of the color symmetry. These identities are going to be
very helpful in the rest of the paper. We outline the general
derivation of the Bethe-Salpeter equations for the diquark states in
Sec.~\ref{Deriv-BS-eq}. The detailed analysis of the Bethe-Salpeter
equations for the NG bosons and the massive diquarks is presented in
Secs.~\ref{BS-eq-NG} and \ref{BS-eq-mass}, respectively.
Appendix~\ref{Ang-int} contains some useful formulas that we use 
through the paper. In Appendix~\ref{non-pert}, we estimate the effect
of the correction to the Schwinger-Dyson equation that comes from the
non-perturbative contribution to the vertex function. At last, in
Appendix~\ref{diffequation}, we present the approximate analytical
solutions to the BS equations.

\section{The model and notation}
\label{model}

In the case of two flavor dense QCD, the original gauge symmetry 
$SU(3)_{c}$  breaks down to the $SU(2)_{c}$ by the Higgs mechanism.
The flavor $SU(2)_{L} \times SU(2)_{R}$ group remains intact. The
appropriate order parameter is given by the vacuum expectation value
of the diquark (antidiquark) field that is an antitriplet (triplet) in
color and a singlet in flavor. Without loss of generality, we assume
that the order parameter points in the third direction of the color
space,
\be
\varphi =\langle 0| \varepsilon^{ij} \varepsilon_{3ab} 
\left(\bar{\Psi}_{D}\right)^{a}_{i}
\gamma^{5} \left(\Psi_{D}^{C}\right)^{b}_{j} |0\rangle ,
\label{order-par}
\ee
where $\Psi_{D}$ and $\Psi_{D}^{C}=C\bar{\Psi}^{T}_{D}$ are the Dirac
and its charge conjugate spinors, and $C$ is a unitary matrix that
satisfies $C^{-1} \gamma_{\mu} C=-\gamma^{T}_{\mu} $ and $C=-C^{T}$.
Here and in what follows, we explicitly display the flavor
($i,j=1,2$) and color ($a,b=1,2$) indices of the spinor fields. It is
also appropriate to mention that the subscript and superscript indices
correspond to complex conjugate representations. 

The order parameter in Eq.~(\ref{order-par}) is even under parity.
Such a choice is dictated by the instanton induced interactions
\cite{W1,S1} which, despite being vanishingly small at large chemical
potential, could be sufficiently strong for picking up the right
vacuum. In addition, any bare Dirac masses of quarks (which are
non-zero in nature) should also favor the parity-even condensate
\cite{EHS-par,PR-par}. 

With the choice of the order parameter orientation as in 
Eq.~(\ref{order-par}), it is very convenient to introduce the
following Majorana spinors:
\begin{eqnarray}
\Psi^{i}_{a} &=& \psi^{i}_{a}
+ \varepsilon_{3ab} \varepsilon^{ij} (\psi^{C})_{j}^{b} ,
\quad a=1,2,  \label{Maj-psi}\\
\Phi^{i}_{a}  &=& \phi^{i}_{a}
- \varepsilon_{3ab} \varepsilon^{ij} (\phi^{C})_{j}^{b} ,
\quad a=1,2  ,\label{Maj-phi}
\end{eqnarray}
which are built of the Weyl spinors of the first two colors,
\begin{eqnarray}
\psi^{i}_{a}={\cal P}_{+} (\Psi_{D})^{i}_{a}, &\quad &
(\psi^{C})_{j}^{b}={\cal P}_{-} (\Psi_{D}^{C})_{j}^{b}, 
\label{def-psi} \\
\phi^{i}_{a}={\cal P}_{-} (\Psi_{D})^{i}_{a}, &\quad &
(\phi^{C})_{j}^{b}={\cal P}_{+} (\Psi_{D}^{C})_{j}^{b}.
\label{def-phi} 
\end{eqnarray}
Here ${\cal P}_{\pm}=(1 \pm \gamma^5)/2$ are the left- and
right-handed projectors.  The new spinors in Eqs.~(\ref{Maj-psi})
and (\ref{Maj-phi}), as is easy to check from their definition,
satisfy the following generalized Majorana conditions: 
\begin{eqnarray}
\left(\Psi^{C}\right)_{i}^{a}
&=& \varepsilon^{3ab} \varepsilon_{ij} \Psi^{j}_{b} ,
\label{psi-self}\\
\left(\Phi^{C}\right)_{i}^{a}
&=& -\varepsilon^{3ab} \varepsilon_{ij} \Phi^{j}_{b}.
\label{phi-self}
\end{eqnarray}
In the color superconducting phase of QCD in which quarks are known
to acquire a dynamical (Majorana) mass, the use of the four-component
Majorana spinors, built of the Weyl spinors, is most natural. Of
course, when quarks are massive and the chiral symmetry is explicitly
broken, it would be more appropriate to consider the eight-component
Majorana spinors, made of Dirac ones.

With our choice of the order parameter that points in the third
direction of the color space, only the quarks of the first two
colors take part in the condensation. The quarks of the third color
do not participate in the color condensate. It is more convenient,
therefore, to use the left and right Weyl spinors,
\begin{eqnarray}
\psi^{i}={\cal P}_{+} (\Psi_{D})^{i}_{3}, &\quad &
(\psi^{C})_{j}={\cal P}_{-} (\Psi_{D}^{C})_{j}^{3}, 
\label{def-psi-3} \\
\phi^{i}={\cal P}_{-} (\Psi_{D})^{i}_{3}, &\quad &
(\phi^{C})_{j}={\cal P}_{+} (\Psi_{D}^{C})_{j}^{3},
\label{def-phi-3} 
\end{eqnarray}
for their description. Notice that the color index ``3" is omitted
in the definition of $\psi^{i}$ and $\phi^{i}$.

In the color superconducting phase with a parity even condensate
(\ref{order-par}), parity is a good symmetry. Then, all the quantum
states of the Hilbert space, including those in the diquark channel,
could be chosen so that they are either parity-even or parity-odd.
In order to construct such states explicitly, we  would need to know
the following parity transformation properties of the spinors:
\ba
\psi^{i}(x) \to \gamma^0 \phi^{i}(x^{\prime}) ,   &\qquad&
\phi^{i}(x) \to \gamma^0 \psi^{i}(x^{\prime})     ,  \\
\psi^{C}_{j}(x) \to -\gamma^0 \phi^{C}_{j}(x^{\prime}), &\qquad&
\phi^{C}_{j}(x) \to -\gamma^0 \psi^{C}_{j}(x^{\prime})   , \\
\Psi^{i}_{a}(x) \to  \gamma^0 \Phi^{i}_{a}(x^{\prime}) ,  &\qquad&
\Phi^{i}_{a}(x) \to \gamma^0 \Psi^{i}_{a}(x^{\prime})  .
\label{9c}
\ea
\label{parity-tran}
where $x=(x_{0},\vec{x})$ and $x^{\prime}=(x_{0},-\vec{x})$. 

Before concluding this section, let us rewrite the order parameter
(\ref{order-par}) in terms of the Majorana spinors, 
\be
\varphi = -\langle 0| 
\bar{\Psi}^{a}_{i} {\cal P}_{-} \Psi^{i}_{a} 
+\bar{\Phi}^{a}_{i} {\cal P}_{+} \Phi^{i}_{a} |0\rangle .
\label{o-p}
\ee
This representation is explicitly $SU(2)_{L}\times SU(2)_{R} \times
SU(2)_{c}$ invariant, and so it is very convenient. By making use of
the transformation properties in Eq.~(\ref{9c}), we also easily check
that $\varphi$ is even under parity, as it should be.

\section{Schwinger-Dyson equation}
\label{SD-equation}

In order to have a self-contained discussion, in this section we
briefly review the Schwinger-Dyson (SD) equation using our new
notations. This would also serve us as a convenient reference point
when we discuss more complicated Bethe-Salpeter (BS) equations.

To start with, let us introduce the following multi-component spinor:
\be
\left(\begin{array}{c}
\Psi^{i}_{a} \\
\psi^{i} \\
\psi^{C}_{i}
\end{array}\right), 
\label{multi-com}
\ee
built of the left fields alone. Similarly, we could introduce a
multi-spinor made of the right fields. In our analysis, restricted
only to the (hard dense loop improved) rainbow approximation, the left
and right sectors of the theory completely decouple. Then, without
loss of generality, it is sufficient to study the SD equation only in
one of the sectors. 

With the notation in Eq.~(\ref{multi-com}), the inverse full
propagator of quarks takes a particularly simple block-diagonal form,
\be
G^{-1}_{p}=\mbox{diag} 
\left(S^{-1}_{p}\delta_{a}^{~b}\delta^{i}_{~j}~,~
s^{-1}_{p}\delta^{i}_{~j}~,~
\bar{s}^{-1}_{p}\delta_{i}^{~j}\right),
\label{propagator}
\ee
where, upon neglecting the wave functions renormalization effects
of quarks \cite{Son,us,SW2,PR2,H1,Brown1,us2},
\begin{eqnarray}
S^{-1}_{p} &=& 
-i\left( \not{\! p} +\mu \gamma^0 \gamma^5 + \Delta_{p} {\cal P}_{-}
+ \tilde{\Delta}_{p} {\cal P}_{+} \right) \nonumber \\
&=& -i \left[ \left(p_{0} -\epsilon^{-}_{p} \right)
\gamma^{0} \Lambda^{+}_{p}
+(\Delta^{-}_{p} )^{*} \Lambda^{+}_{p}
+\left(p_{0} +\epsilon^{+}_{p}\right)
\gamma^{0} \Lambda^{-}_{p}
+(\Delta^{+}_{p} )^{*} \Lambda^{-}_{p}
\right] {\cal P}_{+} \nonumber \\
&& -i \left[ \left(p_{0} -\epsilon^{+}_{p}\right)
\gamma^{0} \Lambda^{+}_{p} +\Delta^{+}_{p} \Lambda^{+}_{p}
+\left(p_{0} +\epsilon^{-}_{p}\right)
\gamma^{0} \Lambda^{-}_{p} +\Delta^{-}_{p} \Lambda^{-}_{p}
\right] {\cal P}_{-} , \label{S_L-inv} \\
s^{-1}_{p}  &=& -i \left( \not{\! p} + \mu \gamma^0  \right) 
{\cal P}_{+} = -i \gamma^{0} \left[
\left(p_{0} -\epsilon^{-}_{p} \right) \Lambda^{+}_{p}
+\left(p_{0} +\epsilon^{+}_{p} \right) \Lambda^{-}_{p}
\right] {\cal P}_{+} , \label{s_L-inv} \\
\bar{s}^{-1}_{p}  &=& -i \left( \not{\! p} - \mu \gamma^0 \right) 
{\cal P}_{-}= -i \gamma^{0} \left[
\left(p_{0} -\epsilon^{+}_{p} \right) \Lambda^{+}_{p}
+\left(p_{0} +\epsilon^{-}_{p} \right) \Lambda^{-}_{p}
\right] {\cal P}_{-} , \label{s^C_L-inv} 
\end{eqnarray}
with $\epsilon^{\pm }_{p}=|\vec{p}|\pm \mu $. The notation for the gap
function, $\Delta_{p} = \Delta^{+}_{p} \Lambda^{+}_{p}  +
\Delta^{-}_{p}\Lambda^{-}_{p}$ and  $\tilde{\Delta}_{p}=\gamma^0
\Delta^{\dagger}_{p}\gamma^0$, as well as the ``on-shell" 
projectors of quarks 
\ba
\Lambda_{p}^{\pm }=\frac{1}{2}
\left(1\pm \frac{\vec{\alpha} \cdot \vec{p}}{|\vec{p}|}\right),
\quad \vec{\alpha} =\gamma^{0} \vec{\gamma} ,
\ea
are the same as in Ref.~\cite{us}. 

\begin{figure}
\epsfbox[50 265 400 365]{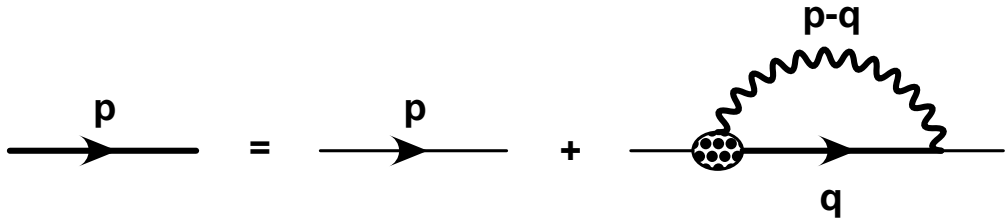}
\caption{The diagrammatic representation of the SD equation.}
\label{fig-sd-eq}
\end{figure}

Now, it is straightforward to derive the matrix form of the SD
equation, 
\be
G^{-1}_{p}=\left(G^{0}_{p}\right)^{-1}
+4\pi\alpha_{s}\int\frac{d^4 q}{(2\pi)^4}
\gamma^{A\mu} G_{q} \Gamma^{A\nu}(q,p)
{\cal D}_{\mu\nu}(q-p),
\label{SD-eq}
\ee
where $\gamma^{A\mu}$ and $\Gamma^{A\mu}$ are the bare and the full
vertices, respectively. This equation is diagrammatically presented in
Fig.~\ref{fig-sd-eq}. The thin and bold solid lines correspond to the
bare and full quark propagators, respectively.  The wavy line stands
for the full gluon propagator.


The only complication of using the multi-component spinor
(\ref{multi-com}) appears due to a more involved structure of the
quark-gluon interaction vertex. Indeed, the explicit form of the bare
vertex reads
\be
\gamma^{A\mu}=\gamma^{\mu}\left(\begin{array}{ccc}
\left[ (T^{A})_{a}^{~b}-2\delta^{A}_{8} (T^{8})_{a}^{~b}
{\cal P}_{-}\right] \delta^{i}_{~j} &
(T^{A})_{a}^{~3} {\cal P}_{+} \delta^{i}_{~j}  &
-\hat{\varepsilon}_{ac}^{ij} (T^{A})_{3}^{~c} {\cal P}_{-} \\
(T^{A})_{3}^{~b} {\cal P}_{+} \delta^{i}_{~j} &
(T^{A})_{3}^{~3} {\cal P}_{+} \delta^{i}_{~j} & 0 \\
-(T^{A})_{c}^{~3} \hat{\varepsilon}^{cb}_{ij} {\cal P}_{-}  &
0 & -(T^{A})_{3}^{~3} {\cal P}_{-} \delta_{i}^{~j}
\end{array}\right),
\label{vertex}
\ee
where $\hat{\varepsilon}_{ac}^{ij} \equiv \varepsilon^{ij}
\varepsilon_{3ac}$ and $T^{A}$ are the $SU(3)_{c}$ generators in the
fundamental representation [$\mbox{tr} (T^{A}T^{B}) = \delta^{AB}/2$].
The relatively complicated structure of the bare vertex might suggest
that our notation are somewhat unnatural. As we shall see in
Sec.~\ref{Ward-id},  because of the breakdown of the $SU(3)_{c}$
symmetry, this structure, on the contrary, is quite natural, and it is
especially so in the case of the full quark-gluon vertex function.

The gluon propagator in the SD equation is the same as in
Ref.~\cite{us}. When the Meissner effect is neglected, the propagator
in the Euclidean space ($k_0=i k_4$) reads
\ba
{\cal D}_{\mu\nu}^{AB}(ik_4,\vec{k}) &\equiv & 
\delta^{AB} {\cal D}_{\mu\nu}(ik_4,\vec{k}) 
\simeq 
i\delta^{AB} \frac{|\vec{k}|}{|\vec{k}|^3+\pi M^2 |k_4|/2}
O^{(1)}_{\mu\nu} \nonumber \\
&& +i\delta^{AB} \frac{1}{k_4^2+|\vec{k}|^2+2 M^2 }
O^{(2)}_{\mu\nu}
+i\delta^{AB} \frac{ d }{k_4^2+|\vec{k}|^2} O^{(3)}_{\mu\nu},
\label{D-long}
\ea
where $M^2=N_{f}\alpha_{s}\mu^2/\pi$ (with $N_{f}=2$), and 
$O^{(i)}_{\mu\nu}$ are the projection operators of three different
types of gluons (magnetic, electric and longitudinal, respectively),
see Ref.~ \cite{us}. The Meissner effect could be qualitatively taken
into account by the following replacement of the magnetic term
\cite{us}:
\be
i\delta^{AB} \frac{|\vec{k}|}{|\vec{k}|^3+\pi M^2 |k_4|/2}
O^{(1)}_{\mu\nu} \quad \to \quad
i\delta^{AB} \frac{|\vec{k}|}{|\vec{k}|^3
+\pi M^2 \left(|k_4|+ c |\Delta^{-}_{0}| \right)/2} 
O^{(1)}_{\mu\nu} 
\label{Meissner}
\ee
in the propagators of those five gluons that correspond to the broken
color generators ($A,B=4,\ldots,8$). In this last expression, $c=O(1)$
is a constant of order one.

By inverting the expression in Eq.~(\ref{propagator}), we obtain the
following representation for the quark propagator:
\ba
G_{p} &=& \mbox{diag} \left(
S_{p} \delta_{a}^{~b} \delta^{i}_{~j}~,~
s_{p} \delta^{i}_{~j}~,~
\bar{s}_{p} \delta_{i}^{~j}
\right), \label{propagator-L}\\
S_{p} &=&
i\frac{\gamma^0 (p_0+\epsilon_{p}^{+})-\Delta^{+}_{p} }
{p_0^2-(\epsilon_{p}^{+})^2-|\Delta^{+}_{p} |^2}
\Lambda_{p}^{-} {\cal P}_{+}
+i\frac{\gamma^0 (p_0-\epsilon_{p}^{+})-(\Delta^{+}_{p} )^{*}}
{p_0^2-(\epsilon_{p}^{+})^2-|\Delta^{+}_{p} |^2}
\Lambda_{p}^{+} {\cal P}_{-} \nonumber \\
&+& i\frac{\gamma^0 (p_0-\epsilon_{p}^{-})-\Delta^{-}_{p} }
{p_0^2-(\epsilon_{p}^{-})^2-|\Delta^{-}_{p} |^2}
\Lambda_{p}^{+} {\cal P}_{+}
+i\frac{\gamma^0 (p_0+\epsilon_{p}^{-})-(\Delta^{-}_{p} )^{*}}
{p_0^2-(\epsilon_{p}^{-})^2-|\Delta^{-}_{p} |^2}
\Lambda_{p}^{-} {\cal P}_{-} , \label{S_L} \\
s_{p}  &=& i\frac{\gamma^0 \Lambda_{p}^{+}{\cal P}_{-}}
{p_0+\epsilon_{p}^{+}} +i\frac{\gamma^0 \Lambda_{p}^{-}{\cal P}_{-}}
{p_0-\epsilon_{p}^{-}},\label{s_L} \\
\bar{s}_{p}  &\equiv&  C s^{T}_{-p} C^{\dagger}
=i\frac{\gamma^0 \Lambda_{p}^{+}{\cal P}_{+}}
{p_0+\epsilon_{p}^{-}} +i\frac{\gamma^0 \Lambda_{p}^{-}{\cal P}_{+}}
{p_0-\epsilon_{p}^{+}} . \label{s^C_L} 
\ea
The bare propagator in Eq.~(\ref{SD-eq}) is similar but with zero
value of the gap. 

In the improved ladder (rainbow) approximation, both vertices in the
SD equation are bare. By making use of the propagators in
Eqs.~(\ref{D-long}) and (\ref{propagator-L}), along with the vertex
in Eq.~(\ref{vertex}), we derive the well known gap equation
\cite{us,SW2,PR2,H1,Brown1,us2},
\be
\Delta^{-}_{p} = \frac{4}{3}\pi\alpha_{s}\int\frac{d^4 q}{(2\pi)^4}
\frac{\Delta^{-}_{q} \mbox{tr} \left( \gamma^{\mu} \Lambda^{+}_{q} 
\gamma^{\nu} \Lambda^{-}_{p} \right) }
{q_0^2-(\epsilon_{q}^{-})^2-|\Delta^{-}_{q}|^2} 
{\cal D}_{\mu\nu}(q-p).
\label{gap-eq}
\ee
After calculating the trace and performing the angular integration
(see Appendix~\ref{Ang-int}), this equation considerably simplifies. 
Then, by assuming that the dependence of the gap on the spatial
component of the momentum is irrelevant in the vicinity of the Fermi
surface, one arrives at the following approximate equation
\cite{Son,us,SW2,PR2,H1,Brown1}:
\begin{equation}
\Delta^{-} (p_{4})\simeq\frac{2\alpha_{s}}{9\pi}
\int_{0}^{\Lambda}  \frac{d q_{4} \Delta^{-} (q_{4})}
{\sqrt{q_{4}^2+|\Delta^{-}_{0}|^2}}
\ln\frac{\Lambda}{|p_{4}-q_{4}|},
\label{sd-eq-appr}
\end{equation}
where  $\Lambda = (4\pi)^{3/2} \mu/\alpha_{s}^{5/2}$. The analytical
solution to this equation is relatively easy to obtain 
\cite{us},
\begin{mathletters}
\ba 
\Delta^{-} (p_{4}) &\simeq& |\Delta^{-}_{0}| J_{0}
\left(\nu\sqrt{\frac{p_{4}}{|\Delta^{-}_{0}|}}\right),
\quad p_{4} \leq |\Delta^{-}_{0}| , \\
\Delta^{-} (p_{4}) &\simeq& |\Delta^{-}_{0}| 
\sqrt{J_{0}^{2}(\nu)+J_{1}^{2}(\nu)} \sin
\left(\frac{\nu}{2} \ln \frac{\Lambda}{p_{4}}\right),
\quad p_{4} \geq |\Delta^{-}_{0}| ,
\ea
\end{mathletters}
where $J_{i}(z)$ are the Bessel functions and $\nu=
\sqrt{8\alpha_{s}/9\pi}$. The corresponding result for the value of
the gap reads
\be
|\Delta^{-}_{0}|\simeq \frac{(4\pi)^{3/2}\mbox{e} \mu}
{\alpha_{s}^{5/2}}
\exp\left(- \frac{3 \pi^{3/2}}{2^{3/2}\sqrt{\alpha_{s}}}
\right),
\label{A-d}
\ee
where $\mbox{e}=2.718\ldots$. Most of the existing studies
\cite{Son,us,SW2,PR2,H1,Brown1,us2} seem to agree upon the dependence
of this result on the coupling constant. The issue of the overall
constant factor, however, is still not settled  down. The analysis of
Ref.~\cite{Brown1}, for example, suggests  that the wave function
renormalization effects of quarks give an extra factor of order
one\footnote{Note, that the argument of Ref.~\cite{Brown1} is somewhat
incomplete, since the calculation is performed for the critical
temperature rather than the order parameter itself. The celebrated BCS
relation between the critical temperature and the gap might not be
satisfied after the Meissner effect is carefully taken into account.}.
Another source of corrections might be due to the running of the
coupling constant \cite{0004013}. In addition to these, as we argue in
Appendix~\ref{non-pert} using the Ward identities, there also exists
at least one non-perturbative correction that could modify the
constant factor in the expression for the gap.

\section{Ward identities}
\label{Ward-id}

As in any other gauge theory, in order to preserve the gauge
invariance in QCD, one has to make sure that some exact relations
(Ward identities) between Green functions are satisfied. In this
section, we consider the simplest Ward identities that relate the
vertex functions and the quark propagators. In addition to
establishing the longitudinal part of the full vertex function, these
identities will play a very important role in our analysis of the BS
equations for the NG and pseudo-NG bosons. 

In general, the structure of Ward identities in non-Abelian gauge
theories (the Slavnov-Taylor identities) is much more complicated than
in Abelian ones: they include contributions of the  Faddeev-Popov
ghosts. Fortunately in the (hard dense loop improved) ladder
approximation, used in this paper, the situation simplifies. Indeed,
since the direct interactions between gluons are neglected in this
approximation, the Ward identities have an Abelian-like structure.

To start with, let us rewrite the conserved currents (related to the
color symmetry) in terms of the Majorana fields, defined in
Eqs.~(\ref{Maj-psi}) and (\ref{Maj-phi}), and the Weyl spinors of  the
third color. By making use of their definitions, it is straightforward
to obtain the following representation for the currents:  
\ba
j^{A}_{\mu}(x) &=& \bar{\Psi}_{D}(x) \gamma_{\mu} T^{A} \Psi_{D}(x)
\nonumber\\ &=& 
\frac{1}{2\sqrt{3}} \delta^{A}_{8} \bar{\Psi}^{a}_{i} (x) 
  \gamma_{\mu} {\cal P}_{+} \Psi_{a}^{i}(x) 
+\bar{\psi}_{i} (x) \gamma_{\mu} T^{Aa}_{3} 
  {\cal P}_{+} \Psi_{a}^{i}(x)
+\bar{\Psi}^{a}_{i} (x) {\cal P}_{-} \gamma_{\mu} 
  T^{A3}_{a} \psi^{i}(x)
+\bar{\psi}_{i} (x) \gamma_{\mu} T^{A3}_{3} \psi^{i}(x) 
  \nonumber\\
&+&\frac{1}{2\sqrt{3}} \delta^{A}_{8} \bar{\Phi}^{a}_{i} (x) 
  \gamma_{\mu} {\cal P}_{-} \Phi_{a}^{i}(x)
+\bar{\phi}_{i} (x) \gamma_{\mu} T^{Aa}_{3} 
  {\cal P}_{-} \Phi_{a}^{i}(x)
+\bar{\Phi}^{a}_{i} (x) {\cal P}_{+} \gamma_{\mu} 
  T^{A3}_{a} \phi^{i}(x)
+\bar{\phi}_{i} (x) \gamma_{\mu} T^{A3}_{3}  \phi^{i}(x),
\label{current}
\ea
where, in agreement with the remark made above, the irrelevant (in
this approximation) contributions of gluons and ghosts are omitted.
Here and in the rest of this section, we assume that $A=4,\ldots,8$,
i.e., we do not consider the currents which correspond to the
generators of the unbroken $SU(2)_{c}$ subgroup. 

As we mentioned earlier, we are interested in the Ward identities
that relate the quark-gluon vertices to the propagators of quarks. 
Therefore, let us introduce the following (non-amputated) vertex
functions:
\begin{mathletters}
\ba 
\bfGamma^{A,i}_{aj,\mu}(x,y) &=& \langle 0 | T 
j^{A}_{\mu}(0) \Psi^{i}_{a}(x) \bar{\psi}_{j}(y) 
| 0 \rangle , \label{cur-2} \\
\bfGamma^{A,ai}_{j,\mu}(x,y) &=& \langle 0 | T 
j^{A}_{\mu}(0) \psi^{i}(x) \bar{\Psi}^{a}_{j}(y) 
| 0 \rangle , \label{cur-1}  \\
\bfGamma^{A,i}_{j,\mu}(x,y) &=& \langle 0 | T 
j^{A}_{\mu}(0) \Psi^{i}_{a}(x) \bar{\Psi}^{a}_{j}(y) 
| 0 \rangle , \label{cur-3} \\
\bftGamma^{A,i}_{aj,\mu}(x,y) &=& \langle 0 | T 
j^{A}_{\mu}(0) \Phi^{i}_{a}(x) \bar{\phi}_{j}(y) 
| 0 \rangle , \label{cur-5} \\
\bftGamma^{A,ai}_{j,\mu}(x,y) &=& \langle 0 | T 
j^{A}_{\mu}(0) \phi^{i}(x) \bar{\Phi}^{a}_{j}(y) 
| 0 \rangle , \label{cur-4} \\
\bftGamma^{A,i}_{j,\mu}(x,y) &=& \langle 0 | T 
j^{A}_{\mu}(0) \Phi^{i}_{a}(x) \bar{\Phi}^{a}_{j}(y) 
| 0 \rangle . \label{cur-6}
\ea
\label{vertices}
\end{mathletters}
Besides the operator of the conserved current, the first three
vertices include only the left-handed quark fields, while the other
three vertices contain only the right-handed fields. Because of the
invariance under the parity, all the mixed, left-right vertices are
trivial. For this reason, they are of no special interest here, and we
do not consider them. 

As usual, in order to derive the Ward identities, one needs to know
the transformation properties of the quark fields. The color symmetry
transformations of the Dirac spinors are well known. By making use
of them, it is straightforward to derive the following infinitesimal
transformations for the spinors of interest:
\ba 
\delta \psi^{i} &=& i \omega^{A} \left(
T^{Aa}_{3} {\cal P}_{+} \Psi^{i}_{a} 
+ T^{A3}_{3} \psi^{i} \right) ,\label{del-psi} \\
\delta \bar{\psi}_{i} &=&  -i \omega^{A} \left(
\bar{\Psi}_{i}^{a} {\cal P}_{-} T^{A3}_{a}  
+ \bar{\psi}_{i} T^{A3}_{3} \right) ,\label{del-psi-bar} \\
\delta \Psi_{a}^{i} &=&  i \omega^{A} \left(
T^{Ab}_{a} {\cal P}_{+} \Psi^{i}_{b} +T^{A3}_{a} \psi^{i}
-\varepsilon_{3ab} \varepsilon^{ij} T^{Ab}_{3} \psi^{C}_{j}
-\delta^{A}_{8}\frac{1}{2\sqrt{3}} {\cal P}_{-} \Psi_{a}^{i}
\right) , \label{del-Psi} \\
\delta \bar{\Psi}^{a}_{i} &=&  -i \omega^{A} \left(
\bar{\Psi}_{i}^{b} {\cal P}_{-} T^{Aa}_{b} + \bar{\psi}_{i} T^{Aa}_{3} 
-\varepsilon^{3ab} \varepsilon_{ij} \bar{\psi}^{Cj} T^{A3}_{b} 
-\delta^{A}_{8}\frac{1}{2\sqrt{3}}\bar{\Psi}^{a}_{i} {\cal P}_{+}
\right) , \label{del-Psi-bar} 
\ea
for the left-handed fields, and
\ba 
\delta \phi^{i} &=& i \omega^{A} \left(
T^{Aa}_{3} {\cal P}_{-} \Phi^{i}_{a} 
+ T^{A3}_{3} \phi^{i} \right) ,\label{del-phi} \\
\delta \bar{\phi}_{i} &=&  -i \omega^{A} \left(
\bar{\Phi}_{i}^{a} {\cal P}_{+} T^{A3}_{a}  
+ \bar{\phi}_{i} T^{A3}_{3} \right) ,\label{del-phi-bar} \\
\delta \Phi_{a}^{i} &=&  i \omega^{A} \left(
T^{Ab}_{a} {\cal P}_{-} \Phi^{i}_{b} +T^{A3}_{a} \phi^{i}
-\varepsilon_{3ab} \varepsilon^{ij} T^{Ab}_{3} \phi^{C}_{j}
-\delta^{A}_{8}\frac{1}{2\sqrt{3}} {\cal P}_{+} \Phi_{a}^{i}
\right) , \label{del-Phi} \\
\delta \bar{\Phi}^{a}_{i} &=&  -i \omega^{A} \left(
\bar{\Phi}_{i}^{b} {\cal P}_{+} T^{Aa}_{b} + \bar{\phi}_{i} T^{Aa}_{3} 
-\varepsilon^{3ab} \varepsilon_{ij} \bar{\phi}^{Cj} T^{A3}_{b} 
-\delta^{A}_{8}\frac{1}{2\sqrt{3}}\bar{\Phi}^{a}_{i} {\cal P}_{-}
\right) , \label{del-Phi-bar} 
\ea
for the right-handed fields. In all these expressions, $\omega^{A}$
are small parameters, parameterizing the transformations of the
$SU(3)_{c}$ group.

In a standard way, by making use of the current conservation as well
as the definition of the vertices in Eq.~(\ref{vertices}), we 
obtain the following Ward identities for the non-amputated vertices:
\begin{mathletters}
\ba
P^{\mu} \bfGamma^{A,i}_{aj,\mu}(k+P,k) &=&i T^{A3}_{a}  
\delta^{i}_{j} \left[ s_{k} -S_{k+P} \right] {\cal P}_{-},\label{27b} \\
P^{\mu} \bfGamma^{A,ai}_{j,\mu}(k+P,k) &=&
i T^{Aa}_{3} \delta^{i}_{j} 
{\cal P}_{+} \left[ S_{k} -s_{k+P} \right], \label{27a}\\
P^{\mu} \bfGamma^{A,i}_{j,\mu}(k+P,k) &=&
\frac{i}{2\sqrt{3}} \delta^{A}_{8} \delta^{i}_{j}
\left[ 2{\cal P}_{+} S_{k}-{\cal P}_{-} S_{k} 
- 2S_{k+P}{\cal P}_{-}+S_{k+P}{\cal P}_{+}\right],\label{27c} \\
P^{\mu} \bftGamma^{A,i}_{aj,\mu}(k+P,k) &=&i T^{A3}_{a}  
\delta^{i}_{j} \left[ \tilde{s}_{k} -\tilde{S}_{k+P} \right] 
{\cal P}_{+}, \\
P^{\mu} \bftGamma^{A,ai}_{j,\mu}(k+P,k) &=&
i T^{Aa}_{3} \delta^{i}_{j} 
{\cal P}_{-} \left[ \tilde{S}_{k} -\tilde{s}_{k+P} \right], \\
P^{\mu} \bftGamma^{A,i}_{j,\mu}(k+P,k) &=&
\frac{i}{2\sqrt{3}} \delta^{A}_{8} \delta^{i}_{j}
\left[ 2{\cal P}_{-} \tilde{S}_{k}- {\cal P}_{+} \tilde{S}_{k} 
- 2\tilde{S}_{k+P}{\cal P}_{+}+\tilde{S}_{k+P}{\cal P}_{-}\right],
\ea
\label{Ward-LR}
\end{mathletters}
where $S_{k}$, $s_{k}$ and $\tilde{S}_{k}$, $\tilde{s}_{k}$ are
the Fourier transforms of the quark propagators in the left and
right sectors, respectively,
\ba
S(x-y) \delta^{b}_{a} \delta^{i}_{j} &=& \langle 0 | T 
\Psi^{i}_{a}(x) \bar{\Psi}^{b}_{j}(y) | 0 \rangle , \\
s(x-y) \delta^{i}_{j} &=& \langle 0 | T 
\psi^{i}(x) \bar{\psi}_{j}(y) | 0 \rangle , \\
\tilde{S}(x-y) \delta^{b}_{a} \delta^{i}_{j} &=& \langle 0 | T 
\Phi^{i}_{a}(x) \bar{\Phi}^{b}_{j}(y) | 0 \rangle , \\
\tilde{s}(x-y) \delta^{i}_{j} &=& \langle 0 | T 
\phi^{i}(x) \bar{\phi}_{j}(y) | 0 \rangle .
\ea
As we discussed in Sec.~\ref{SD-equation}, in the approximation with
no wave function renormalization effects, the explicit form of the
momentum space propagators for the left-handed fields is given in
Eqs.~(\ref{S_L}) and (\ref{s_L}).  For the completeness of our
presentation, we also mention that the right-handed propagators are
the same, except that the projectors ${\cal P}_{-}$ and ${\cal P}_{+}$
interchange.

At this point, let us note that the use of the non-amputated vertices
in this section is not accidental. In fact, it is crucial for a quick
derivation of the Ward identities. Other than that, the non-amputated
vertices are not very convenient to work with. In  fact, it is
amputated rather than non-amputated vertices that are usually used in
the Feynman diagrams. For example, both the bare and full vertices in
Fig.~\ref{fig-sd-eq} are the amputated ones. Similarly, it is the
amputated vertices that appear in the BS equation in
Sec.~\ref{Deriv-BS-eq}. The formal definitions of the amputated
vertices read 
\begin{mathletters}
\ba
\Gamma^{A,i}_{aj,\mu}(k+P,k) &=&
  S^{-1}_{k+P} \bfGamma^{A,i}_{aj,\mu}(k+P,k) s^{-1}_{k}, 
\label{30b}\\
\Gamma^{A,ai}_{j,\mu}(k+P,k) &=& 
  s^{-1}_{k+P} \bfGamma^{A,ai}_{j,\mu}(k+P,k) S^{-1}_{k}, 
\label{30a}\\
\Gamma^{A,i}_{j,\mu}(k+P,k) &=&
  S^{-1}_{k+P} \bfGamma^{A,i}_{j,\mu}(k+P,k) S^{-1}_{k},
\label{30c} \\
\tilde{\Gamma}^{A,i}_{aj,\mu}(k+P,k) &=& \tilde{S}^{-1}_{k+P} 
\bftGamma^{A,i}_{aj,\mu}(k+P,k) \tilde{s}^{-1}_{k}, \\
\tilde{\Gamma}^{A,ai}_{j,\mu}(k+P,k) &=&  \tilde{s}^{-1}_{k+P}
\bftGamma^{A,ai}_{j,\mu}(k+P,k)  \tilde{S}^{-1}_{k}, \\
\tilde{\Gamma}^{A,i}_{j,\mu}(k+P,k) &=& \tilde{S}^{-1}_{k+P} 
\bftGamma^{A,i}_{j,\mu}(k+P,k) \tilde{S}^{-1}_{k}.
\ea
\label{amp-def}
\end{mathletters}
These, as is clear from our discussion above, are directly related to
the quark-gluon interactions. As is clear from Eq.~(\ref{Ward-LR}),
they satisfy the following identities of their own:
\begin{mathletters}
\ba
P^{\mu} \Gamma^{A,i}_{aj,\mu}(k+P,k) &=&i T^{A3}_{a} \delta^{i}_{j} 
\left[S^{-1}_{k+P} - s^{-1}_{k}\right] {\cal P}_{+},\label{ab} \\  
P^{\mu} \Gamma^{A,ai}_{j,\mu}(k+P,k) &=&i T^{Aa}_{3} \delta^{i}_{j} 
{\cal P}_{-} \left[s^{-1}_{k+P} -S^{-1}_{k} \right], \label{aa}\\
P^{\mu} \Gamma^{A,i}_{j,\mu}(k+P,k) &=&
\frac{i}{2\sqrt{3}} \delta^{A}_{8} \delta^{i}_{j}
\left[ 2S^{-1}_{k+P} {\cal P}_{+} -S^{-1}_{k+P} {\cal P}_{-} 
- 2{\cal P}_{-} S^{-1}_{k}
+{\cal P}_{+} S^{-1}_{k}\right],\label{ac} \\
P^{\mu} \Gamma^{A,i}_{aj,\mu}(k+P,k) &=&i T^{A3}_{a}  
\delta^{i}_{j} \left[ \tilde{S}^{-1}_{k+P} -\tilde{s}^{-1}_{k}
\right] {\cal P}_{-}, \\
P^{\mu} \Gamma^{A,ai}_{j,\mu}(k+P,k) &=&
i T^{Aa}_{3} \delta^{i}_{j} {\cal P}_{+} 
\left[ \tilde{s}^{-1}_{k+P} -\tilde{S}^{-1}_{k}\right], \\
P^{\mu} \Gamma^{A,i}_{j,\mu}(k+P,k) &=&
\frac{i}{2\sqrt{3}} \delta^{A}_{8} \delta^{i}_{j} \left[ 
2\tilde{S}^{-1}_{k+P}{\cal P}_{-} -\tilde{S}^{-1}_{k+P} {\cal P}_{+}
- 2{\cal P}_{+} \tilde{S}^{-1}_{k}
+{\cal P}_{-} \tilde{S}^{-1}_{k}\right].
\ea
\label{Ward-amp}
\end{mathletters}
In the rest of the paper, we are going to use these Ward identities a
number times. Because of a relatively simple structure of the inverse
quark propagators, this last form of the identities will be
particularly convenient. 

In connection with the Ward identities, it is appropriate to mention
here the complementary analysis of Ref.~\cite{Brown2}. The authors of
that paper consider the contribution to the  Ward identity that is
directly related to the wave function  renormalization of quarks.

\section{Derivation of the BS equation}
\label{Deriv-BS-eq}

In quantum field theory, bound states and resonances reveal themselves
through the appearance of poles in Green functions. These latter
satisfy some general BS equations which usually are rather
complicated. 

To consider the problem of diquark bound states in cold dense QCD, one
has to introduce a four-point Green function that describes the two
particle scattering in the diquark channel of interest. The residue at
the pole of the Green function is related to the BS wave function of
the bound state. By starting from the (inhomogeneous) BS equation for
the four-point Green function, it  is straightforward to derive the
so-called homogeneous BS equation for the wave function. 

In the problem at hand, we could construct quite a few different
diquark states. Not all of them could actually be bound states. For
example, one would not expect from a two particle state to form a
bound state unless there is some attraction in the corresponding
channel. Now, in the dense QCD, the dominant interaction between
quarks is given by the one-gluon exchange. As we know, this
interaction is attractive only in the antisymmetric diquark channels.
Therefore, without loss of generality, it is sufficient to consider
only the following bound states:
\begin{mathletters}
\ba
(2\pi)^4\delta^4 (p_{+}-p_{-}-P)
\bfchi^{(\tilde{b})}_{a} (p,P)
&=& (2\pi)^4\delta^4 (p_{+}-p_{-}-P)
\delta^{\tilde{b}}_{a}\bfchi(p,P) \nonumber \\
 &=& \langle 0| T
\Psi^{i}_{a}(p_{+}) \bar{\psi}_{i}(-p_{-})
|P; \tilde{b} \rangle_{L} , \quad \tilde{b}=1,2  , \label{def-chi} \\
(2\pi)^4\delta^4 (p_{+}-p_{-}-P)
\bflam_{(\tilde{a})}^{~b} (p,P)&=& 
(2\pi)^4\delta^4 (p_{+}-p_{-}-P)
\delta^{b}_{\tilde{a}}\bflam(p,P) \nonumber \\
 &=& \langle 0| T
\psi^{i}(p_{+}) \bar{\Psi}_{i}^{b}(-p_{-})
|P; \tilde{a} \rangle_{L} , \quad \tilde{a}=1,2  , \label{def-lambda}\\
(2\pi)^4\delta^4 (p_{+}-p_{-}-P)
\bfeta (p,P)  &=& \langle 0| T \Psi^{i}_{a}(p_{+})
\bar{\Psi}_{i}^{a}(-p_{-}) |P \rangle_{L} \label{def-eta} ,\\
(2\pi)^4\delta^4 (p_{+}-p_{-}-P)
\bfsig (p,P)  &=& \langle 0| T \psi^{i}(p_{+})
\bar{\psi}_{i}(-p_{-}) |P \rangle_{L}  , \label{def-sigma}
\ea
\label{BS-all-L}
\end{mathletters}
plus the states made out of the right handed fields,
\begin{mathletters}
\ba
(2\pi)^4\delta^4 (p_{+}-p_{-}-P)
\bftchi^{(\tilde{b})}_{a} (p,P) &=&
(2\pi)^4\delta^4 (p_{+}-p_{-}-P) 
\delta^{\tilde{b}}_{a}\bftchi(p,P) \nonumber \\
 &=& \langle 0| T
\Phi^{i}_{a}(p_{+}) \bar{\phi}_{i}(-p_{-})
|P;\tilde{b} \rangle_{R} , \quad \tilde{b}=1,2  , \label{def-t-chi} \\
(2\pi)^4\delta^4 (p_{+}-p_{-}-P)
\bftlam_{(\tilde{a})}^{~b} (p,P) &=&
(2\pi)^4\delta^4 (p_{+}-p_{-}-P)
\delta^{b}_{\tilde{a}}\bftlam(p,P) \nonumber \\
 &=& \langle 0| T
\phi^{i}(p_{+}) \bar{\Phi}_{i}^{b}(-p_{-})
|P; \tilde{a} \rangle_{R} , \quad \tilde{a}=1,2  , 
\label{def-t-lambda} \\
(2\pi)^4\delta^4 (p_{+}-p_{-}-P)
\bfteta (p,P)  &=& \langle 0| T \Phi^{i}_{a}(p_{+})
\bar{\Phi}_{i}^{a}(-p_{-}) |P \rangle_{R}  , 
\label{def-t-eta} \\
(2\pi)^4\delta^4 (p_{+}-p_{-}-P)
\bftsig (p,P)  &=& \langle 0| T \phi^{i}(p_{+})
\bar{\phi}_{i}(-p_{-}) |P \rangle_{R} ,
\label{def-t-sigma}
\ea
\label{BS-all-R}
\end{mathletters}
where $p=(p_{+}+p_{-})/2$ and the quantities on the right hand
side of these equations are defined as the Fourier transforms 
of the corresponding BS wave functions in the coordinate space.
Notice that the analogous states containing the charge conjugate
fields of the third color, $\psi^{C}_{i}$ and $\phi^{C}_{i}$, are not
independent. Because of the property in Eqs.~(\ref{psi-self}) and
(\ref{phi-self}), they are related to those already introduced.

For completeness, let us note that the only other diquark channel that
we do not consider here is a triplet under $SU(2)_{c}$. It is however
clear that the repulsion dominates in such a channel because this
triplet comes from the $SU(3)_{c}$ sextet. Notice that, although one
does not expect the appearance of a $\bfsig$ bound state, we keep the
$\bfsig$ wave function in the analysis. This is because the equations
for the BS wave functions of the two singlets, $\bfsig$ and $\bfeta$,
may not decouple. Notice also that the doublet, antidoublet and
singlets coming from the $SU(3)_{c}$ triplet and antitriplet can mix
with the doublet, antidoublet and singlets coming from the $SU(3)_{c}$
nonet (octet plus singlet). 

Before proceeding further with the analysis of the bound states, let
us recall that parity is not broken in dense QCD with two flavors,
see Sec.~\ref{model}. Then, all the bound states can be chosen in
such a way that they are either parity-even or parity-odd. Clearly,
the states in Eqs.~(\ref{BS-all-L}) and (\ref{BS-all-R}) do not share
this property. In order to fix this, we could have constructed the
following scalars and pseudoscalars, 
\ba
|P;n\rangle_{s} &=& \frac{1}{\sqrt{2}}\bigg(
|P;n\rangle_{L} + |P;n\rangle_{R} \bigg), \label{scalar-def}\\
|P;n\rangle_{p} &=& \frac{1}{\sqrt{2}}\bigg(
|P;n\rangle_{L} - |P;n\rangle_{R} \bigg),\label{pseudo-def}
\ea
where $n$ denotes the appropriate state. 

In our analysis, however, we find it more convenient to work with the
bound states constructed of either left-handed or right-handed fields
separately. This is because, in the (hard dense loop improved) ladder
approximation, the two sectors of the theory stay completely
decoupled. Besides that, the dynamics of the left and right fields are
identical in the approximation used. Under these conditions, the
degeneracy of the left and right sectors is equivalent to the
degeneracy of the parity-even and parity-odd ones. In this way, we
reveal the parity doubling property of the spectrum of bound states in
QCD at asymptotically high density of quark matter.\footnote{Notice
that there are some subtleties in applying this parity doubling
argument to the case of the (pseudo-) NG bound states, see
Sec.~\ref{BS-eq-NG}.}

\subsection{Equations for the non-amputated wave functions}
\label{Deriv}

In order to derive the BS equations, we use the method developed in
Ref.~\cite{GMS} for the case of zero chemical potential (for a review,
see Ref.\cite{FGMS}). To this end, we need to know the quark
propagators and the quark-gluon interactions in the color
superconducting phase. From the analysis of the SD equation (see
Sec.~\ref{SD-equation}), we got the structure of the quark propagator.
We also know that the approximation with no wave function
renormalization effects is quite reliable, at least in the leading
order. By combining these facts together, we arrive at the following
effective Lagrangian of quarks:
\ba
&& {\cal L}_{eff} = \bar{\Psi}_{i}^{a} \left(
\pslash +\mu \gamma^0 \gamma^5 + \Delta {\cal P}_{-}
+ \tilde{\Delta} {\cal P}_{+} \right) \Psi^{i}_{a}
+\bar{\Psi}_{i}^{a} \Aslash^{B} \left[ (T^{B})_{a}^{~b}
-2\delta^{B}_{8} (T^{8})_{a}^{~b}{\cal P}_{-}\right] \Psi^{i}_{b}
+\bar{\psi}_{i} \Aslash_{3}^{~b} {\cal P}_{+} \Psi^{i}_{b}
-\bar{\psi}^{Ci}  \Aslash_{a}^{~3}
\hat{\varepsilon}^{ab}_{ij} {\cal P}_{-} \Psi^{j}_{b}
\nonumber \\
&& + \bar{\psi}_{i}
\left(\pslash +\mu \gamma^0 \right) {\cal P}_{+}\psi^{i}
+ \bar{\psi}^{Ci}
\left(\pslash -\mu \gamma^0 \right) {\cal P}_{-}\psi^{C}_{i}
+ \bar{\psi}_{i} \Aslash_{3}^{~3} {\cal P}_{+} \psi^{i}
- \bar{\psi}^{Ci} \Aslash_{3}^{~3} {\cal P}_{-} \psi^{C}_{i}
+\bar{\Psi}_{i}^{a} \Aslash_{a}^{~3} {\cal P}_{+} \psi^{i}
-\bar{\Psi}_{i}^{a} \hat{\varepsilon}_{ab}^{ij}
\Aslash_{3}^{~b} {\cal P}_{-} \psi^{C}_{j}
\nonumber \\
&& +\bar{\Phi}_{i}^{a} \left( \pslash - \mu \gamma^0 \gamma^5
+ \Delta {\cal P}_{+} + \tilde{\Delta} {\cal P}_{-} \right)
\Phi^{i}_{a}
+\bar{\Phi}_{i}^{a} \Aslash^{B} \left[ (T^{B})_{a}^{~b}
-2 \delta^{B}_{8} (T^{8})_{a}^{~b} {\cal P}_{+}\right] \Phi^{i}_{b}
+\bar{\phi}_{i} \Aslash_{3}^{~b} {\cal P}_{-} \Phi^{i}_{b}
-\bar{\phi}^{Ci}  \Aslash_{a}^{~3}
\hat{\varepsilon}^{ab}_{ij} {\cal P}_{+} \Phi^{j}_{b}
\nonumber \\
&&
 + \bar{\phi}_{i}
\left(\pslash +\mu \gamma^0 \right){\cal P}_{-} \phi^{i}
+ \bar{\phi}^{Ci}
\left(\pslash -\mu \gamma^0 \right){\cal P}_{+} \phi^{C}_{i}
+ \bar{\phi}_{i} \Aslash_{3}^{~3} {\cal P}_{-} \phi^{i}
- \bar{\phi}^{Ci} \Aslash_{3}^{~3} {\cal P}_{+} \phi^{C}_{i}
+\bar{\Phi}_{i}^{a} \Aslash_{a}^{~3} {\cal P}_{-} \phi^{i}
-\bar{\Phi}_{i}^{a} \hat{\varepsilon}_{ab}^{ij}
\Aslash_{3}^{~b} {\cal P}_{+} \phi^{C}_{j} ,
\label{L-eff}
\ea
where, by definition, $\Delta = \Delta^{+}_{p} \Lambda^{+}_{p}  +
\Delta^{-}_{p}\Lambda^{-}_{p}$, $\tilde{\Delta}=\gamma^0
\Delta^{\dagger}\gamma^0$, and $\hat{\varepsilon}^{ab}_{ij}
=\varepsilon^{3ab} \varepsilon_{ij}$. The choice of $\Delta$, as is
easy to check, corresponds to the case of the parity-even Majorana
mass.

The effective Lagrangian in Eq.~(\ref{L-eff}) is the starting point
in derivation of the BS equations for the wave functions introduced in
Eqs.~(\ref{BS-all-L}) and (\ref{BS-all-R}). While using the notation
of the multicomponent spinor in Eq.~(\ref{multi-com}), it is natural
to combine the (left-handed) wave functions of the bound states into
the following matrix:
\be
X(p,P) = \left(\begin{array}{ccc}
\frac{1}{2} \bfeta (p,P) \delta_{a}^{~b} \delta^{i}_{~j}  &
\bfchi^{(\tilde{b})}_{a}(p,P)  \delta^{i}_{~j} &
\hat{\varepsilon}^{ij}_{ac} C \left(
\bflam_{(\tilde{a})}^{~c}(-p,P)  \right)^{T} C^{\dagger} \\
\bflam_{(\tilde{a})}^{~b}(p,P) \delta^{i}_{~j}  &
\bfsig(p,P) \delta^{i}_{~j}  & 0 \\
C \left( \bfchi^{(\tilde{b})}_{c}(-p,P) \right)^{T} C^{\dagger}
\hat{\varepsilon}_{ij}^{cb} & 0 & C \bfsig^{T}(-p,P)  C^{\dagger}
\delta_{i}^{~j}
\end{array}\right),
\label{matrix-BS-wf}
\ee
where we took into account the property of Majorana spinors given in
Eq.~(\ref{psi-self}). We could also introduce a similar matrix wave
function for the right-handed fields. Since, however, in the (hard
dense loop improved) ladder approximation, the left-handed and
right-handed sectors decouple, we study one of them in detail, and
only occasionally refer to the other. 

\begin{figure}
\epsfbox[25 210 400 390]{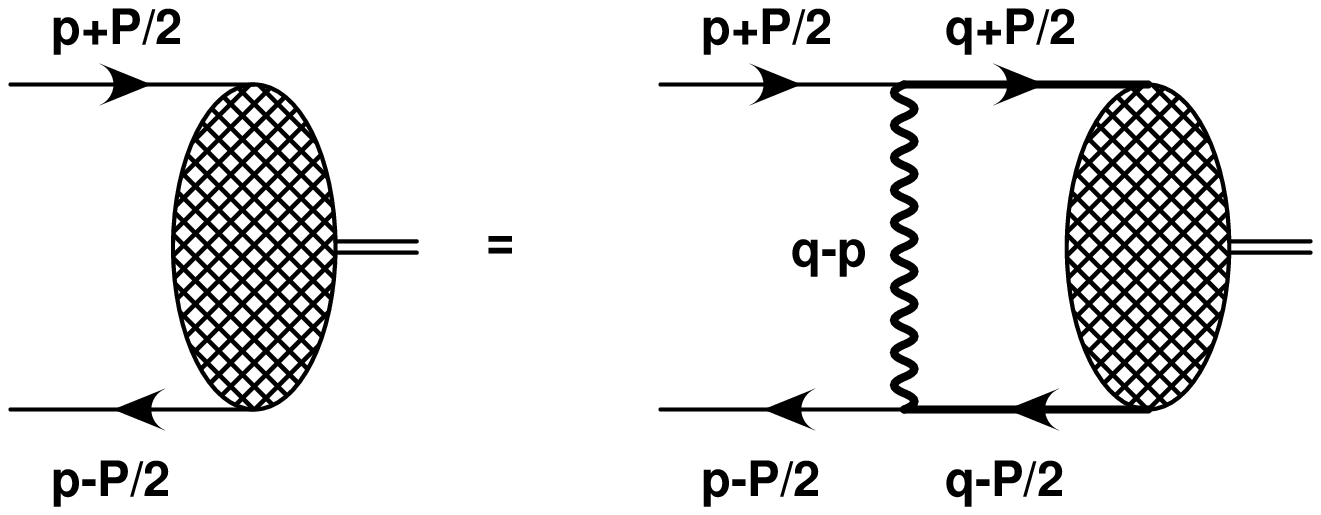}
\caption{The diagrammatic representation of the BS equation.}
\label{fig-bs-eq}
\end{figure}

In the (hard dense loop improved) ladder approximation, the BS wave
function in Eq.~(\ref{matrix-BS-wf}) satisfies the following matrix
equation:
\be
 G^{-1}\left(p+\frac{P}{2}\right)
X(p;P) G^{-1}\left(p-\frac{P}{2}\right)
=-4\pi \alpha_{s}\int\frac{d^4 q}{(2\pi)^4}
\gamma^{A\mu} X(q;P) \gamma^{B\nu}
{\cal D}^{AB}_{\mu\nu}(q-p) , 
\label{BS-matrix-eq}
\ee
where ${\cal D}^{AB}_{\mu\nu}(q-p)$ is the gluon propagator and
$\gamma^{A\mu}$ is the bare quark-gluon vertex. This approximation
has the same status as the rainbow approximation in the SD equation.
It assumes that the coupling constant is weak, and the leading
perturbative expression for the kernel of the BS equation adequately
represents the quark interactions. Schematically, the BS equation
(\ref{BS-matrix-eq}) is shown in Fig.~\ref{fig-bs-eq}. 


By writing it in components, we arrive at the set of four equations,
\ba
S_{p+P/2}^{-1} \bfchi_{a}^{(\tilde{b})} (p,P) s_{p-P/2}^{-1} 
&=& - \frac{2}{3} \pi \alpha_{s} \int
\frac{d^4 q}{(2\pi )^4}  \gamma^{\mu} \left[
{\cal P}_{-} \bfchi^{(\tilde{b})}_{a} (q,P)  {\cal P}_{-}
+3 {\cal P}_{-} C \left(\bfchi^{(\tilde{b})}_{a}
(-q,P)\right)^{T} C^{\dagger} {\cal P}_{-}\right.
\nonumber \\ && \left.
-{\cal P}_{+} \bfchi^{(\tilde{b})}_{a} (q,P)  {\cal P}_{-} \right]
\gamma^{\nu} {\cal D}_{\mu\nu}(q-p), \label{chi} \\
s_{p+P/2}^{-1} \bflam_{(\tilde{a})}^{~b} (p,P) S_{p-P/2}^{-1} 
&=& - \frac{2}{3} \pi \alpha_{s} \int
\frac{d^4 q}{(2\pi )^4} \gamma^{\mu} \left[
{\cal P}_{+} \bflam_{(\tilde{a})}^{~b} (q,P) {\cal P}_{+}
+3 {\cal P}_{+} C \left(
\bflam_{(\tilde{a})}^{~b} (-q,P) \right)^{T} C^{\dagger}
{\cal P}_{+} \right.
\nonumber \\ && \left.
-{\cal P}_{+} \bflam_{(\tilde{a})}^{~b} (q,P) {\cal P}_{-}
\right] \gamma^{\nu} {\cal D}_{\mu\nu}(q-p), \label{lambda}  \\
S_{p+P/2}^{-1} \bfeta (p,P) S_{p-P/2}^{-1}
&=& - \frac{8}{3} \pi \alpha_{s} \int
\frac{d^4 q}{(2\pi )^4} \gamma^{\mu}
\Bigg[ {\cal P}_{+}\bfeta (q,P) {\cal P}_{+}
+ {\cal P}_{-}\bfeta (q,P) {\cal P}_{-}
+ \frac{5}{4} {\cal P}_{-}\bfeta (q,P) {\cal P}_{+}
\nonumber \\ && 
+ \frac{5}{4} {\cal P}_{+}\bfeta (q,P) {\cal P}_{-}
+ \frac{3}{2} {\cal P}_{+} \bfsig (q,P) {\cal P}_{-}
+ \frac{3}{2} {\cal P}_{-} C \bfsig^{T} (-q,P) 
C^{\dagger} {\cal P}_{+}
\Bigg]
\gamma^{\nu} {\cal D}_{\mu\nu}(q-p), \label{eta} \\
s_{p+P/2}^{-1} \bfsig (p,P) s_{p-P/2}^{-1}
&=& - \frac{2}{3} \pi \alpha_{s} \int
\frac{d^4 q}{(2\pi )^4} \gamma^{\mu}
\left[ 3 {\cal P}_{+} \bfeta (q,P) {\cal P}_{-}
+ 2 {\cal P}_{+} \bfsig (q,P) {\cal P}_{-} \right]
\gamma^{\nu} {\cal D}_{\mu\nu}(q-p). \label{sigma} 
\ea
The right-handed fields satisfy a similar set of equations.

In order to solve the BS equations, it is important to determine
the Dirac structure of the BS wave function. There is the following
useful statement. Let us consider a BS wave function of an arbitrary
bound state for a {\em non-zero} chemical potential in the {\em
center of mass frame}. Then, the number of independent terms in its
decomposition over the Dirac matrices coincides with the number of
the terms in the decomposition of the BS wave function at {\em zero}
chemical potential.

The proof of this statement is simple. The Dirac decomposition is
determined by all the space-time tensors characterizing the bound
state, e.g., the momenta $P^{\nu}$ and $p^{\nu}$, the polarization
vector $e^{\nu}$ (in the case of a massive spin one bound state),
etc. In this respect, the case of a non-zero chemical potential is
distinguished by the  occurrence of only one additional vector
$u^{\nu}=(1, \vec{0})$. But in the center of mass frame, where the
total momentum $P^{\nu}=(P_{0}, \vec{0})$, the vector $u^{\nu}$ is
proportional to $P^{\nu}$ and therefore is not independent. Thus,
the number of terms in the Dirac decomposition of a BS wave function
in this frame is the same for both zero and non-zero chemical
potential.

Of course, there is the essential difference between these two
cases: while for zero chemical potential, the number of the Dirac
structures is the same in all the frames, in the case of a non-zero
chemical potential, it is different for $\vec{P}=0$ (the center of
mass frame) and for $\vec{P}\neq 0$ (all other frames). For example,
as we will see for spin zero diquarks, when $\mu\neq 0$, there are
four independent terms in the center of mass frame, and there are
eight terms in other frames.

Strictly speaking, this statement is valid only for massive bound
states. However, in the case of spin zero bound states it is still
valid also for massless states (in particular, for NG bosons): the
point is that the limit $M\to 0$ is smooth for the BS wave functions
of spin zero states, and $P^{\nu}\to 0$ is a very useful limit for
studying properties of the NG bosons.       

In the next two sections we study these BS equations for the diquark
states in detail. In order to approach the problem, we first need to
determine the Dirac structure of the BS wave functions. As we shall
see, the Ward identities, derived in Sec.~\ref{Ward-id}, are  of
great help in dealing with this task. Moreover, in the particular
case of the (pseudo-) NG bosons, the knowledge of the Ward identities
is powerful enough to reveal the complete solution to the BS
equations. We consider this important case in the next section.

\section{BS equations for NG and pseudo-NG bosons}
\label{BS-eq-NG}

In this section we consider the massless bound diquark states. The
latter should include the NG and pseudo-NG bosons. Before proceeding
to the detailed analysis of the BS equations, it is instructive to
describe the qualitative physical picture in the problem at hand.

Let us start from a simple observation. As we stressed many times,
the QCD dynamics at large chemical potential consists of two
essentially decoupled and identical (left-handed and right-handed)
sectors. Then, as long as it concerns the diquark paring dynamics,
no changes would appear in the model if one enlarges the gauge group
of QCD from $SU(3)_{c}$ to the approximate $SU(3)_{c,L}\times
SU(3)_{c,R}$, assuming that the coupling constants of both gauge
groups are identical. In the modified theory, the pattern of the
symmetry breaking should be $SU(3)_{c,L}\times SU(3)_{c,R} \to
SU(2)_{c,L} \times SU(2)_{c,R}$. In this case, ten NG bosons should
appear. If the gauge group were $SU(3)_{c,L}\times SU(3)_{c,R}$, all
the ten NG bosons would be unphysical because of the Higgs
mechanism. However, since the true gauge group of QCD is vector-like
$SU(3)_{c}$, only five NG bosons (scalars) are removed from the
spectrum of physical particles by the Higgs mechanism. The other
five NG bosons (pseudoscalars) should remain in the spectrum. In the
complete theory, these latter are the pseudo-NG bosons. They should
get non-zero masses due to higher order 
corrections that are beyond the improved ladder approximation
(an example of such corrections is the box diagram in the BS
kernel with two intermediate gluons). At
the same time, since the theory is weakly coupled at large chemical
potential, it is natural to expect that the masses of the pseudo-NG
bosons are small even compared to the value  of the dynamical quark
mass \cite{Weinberg}.

For the completeness of our discussion, let us also add that, even
though the massless scalars are removed from the physical spectrum,
they exist in the theory as some kind of ``ghosts" \cite{JJCN}. In
fact, one cannot completely get rid of them, unless a unitary gauge
is found.\footnote{Note that, because of the composite (diquark)
nature of the order parameter in color superconducting phase of
dense QCD, it does not seem to be straightforward to define  and to
use the unitary gauge there.} It is also important to mention that
these ghosts play a very important role in getting rid of
unphysical poles from on-shell scattering amplitudes \cite{JJCN}.

\subsection{The structure of the BS wave functions of 
the (pseudo-) NG bosons}  
\label{BS-str-NG}

Earlier we mentioned in passing that the use of the Ward identities 
is crucial for revealing the Dirac structure of the BS wave functions
of the (pseudo-) NG bosons. Now let us elaborate this point. We start
with the definition of the vertices in Eq.~(\ref{vertices}). By
making use of them, one can show that the corresponding Fourier
transforms develop poles whenever the total momentum of the incoming
quarks, $P$, satisfies the on-shell condition of a bound state. In
particular, as $P\to 0$, we obtain
\begin{mathletters}
\ba 
\left. \bfGamma^{A,i}_{aj,\mu}(p+P/2,p-P/2) \right|_{P\to 0} 
&\simeq & \frac{P^{(\chi)}_{\mu}F^{(\chi)}}{P^{\nu} P^{(\chi)}_{\nu}}
\sum_{\tilde{a}} \delta^{i}_{j} T_{\tilde{a}}^{A3}
\bfchi^{(\tilde{a})}_{a} (p,0) \equiv 
\frac{P^{(\chi)}_{\mu}F^{(\chi)}}{P^{\nu} P^{(\chi)}_{\nu}}
\delta^{i}_{j} T_{a}^{A3}
\bfchi (p,0), \label{con-2} \\
\left. \bfGamma^{A,ai}_{j,\mu}(p+P/2,p-P/2)\right|_{P\to 0} &\simeq& 
\frac{P^{(\lambda)}_{\mu}F^{(\lambda)}}{P^{\nu} P^{(\lambda)}_{\nu}}
\sum_{\tilde{a}} \delta^{i}_{j} T_{3}^{A\tilde{a}}
\bflam^{a}_{(\tilde{a})} (p,0) \equiv 
\frac{P^{(\lambda)}_{\mu}F^{(\lambda)}}{P^{\nu} P^{(\lambda)}_{\nu}}
\delta^{i}_{j} T_{3}^{Aa} \bflam (p,0), 
\label{con-1} \\
\left. \bfGamma^{A,i}_{j,\mu}(p+P/2,p-P/2)\right|_{P\to 0} 
&\simeq & \frac{P^{(\eta)}_{\mu}F^{(\eta)}}{P^{\nu} P^{(\eta)}_{\nu}}
\frac{1}{2} \delta^{i}_{j} \delta_{8}^{A} \bfeta (p,0) , 
\label{con-3} \\
\left. \bftGamma^{A,i}_{aj,\mu}(p+P/2,p-P/2)  \right|_{P\to 0} 
&\simeq & 
\frac{P^{(\chi)}_{\mu}\tilde{F}^{(\chi)}}{P^{\nu} P^{(\chi)}_{\nu}}
\sum_{\tilde{a}} \delta^{i}_{j} T_{\tilde{a}}^{A3}
\bftchi^{(\tilde{a})}_{a} (p,0) \equiv 
\frac{P^{(\chi)}_{\mu}\tilde{F}^{(\chi)}}{P^{\nu} P^{(\chi)}_{\nu}}
\delta^{i}_{j} T_{a}^{A3} \bftchi (p,0) , 
\label{con-5} \\
\left. \bftGamma^{A,ai}_{j,\mu}(p+P/2,p-P/2)  \right|_{P\to 0} 
&\simeq & 
\frac{P^{(\lambda)}_{\mu} 
\tilde{F}^{(\lambda)}}{P^{\nu} P^{(\lambda)}_{\nu}}
\sum_{\tilde{a}} \delta^{i}_{j} T_{3}^{A\tilde{a}}
\bftlam^{a}_{(\tilde{a})} (p,0) \equiv 
\frac{P^{(\lambda)}_{\mu}
\tilde{F}^{(\lambda)}}{P^{\nu} P^{(\lambda)}_{\nu}}
\delta^{i}_{j} T_{3}^{Aa} \bftlam (p,0), 
\label{con-4} \\
\left. \bftGamma^{A,i}_{j,\mu}(p+P/2,p-P/2)  \right|_{P\to 0} 
&\simeq & \frac{P^{(\eta)}_{\mu}\tilde{F}^{(\eta)}}{P^{\nu}
P^{(\eta)}_{\nu}} \frac{1}{2} \delta^{i}_{j} \delta_{8}^{A}
\bfteta (p,0), \label{con-6}
\ea
\label{vertices-con}
\end{mathletters}
where $F^{(x)}$ and $\tilde{F}^{(x)}$ (with $x$ being $\lambda$,
$\chi$ or $\eta$) are the decay constants of the (pseudo-) NG
bosons. The rigorous definition and the calculation of their values
will be given in Subsec.~\ref{decay-consts}. For our purposes here,
it is sufficient to know that they are constants expressed through
the parameters of the theory. Since the Lorentz symmetry is
explicitly broken by the chemical potential, the dispersion relations
of the (pseudo-) NG bosons respect only the spatial rotation
symmetry. In order to take this into account, we introduced the
following four-momentum notation: $P^{(x)}_{\mu} =(P^{0},-c_{x}^{2}
\vec{P})$ where $c_{x}<1$ is the velocity of the appropriate
(pseudo-) NG boson. 

By recalling that the parity is preserved in the color
superconducting phase of dense QCD, we conclude that the decay
constants of the left- and right-handed composites should be equal,
i.e., $\tilde{F}^{(\lambda)} =F^{(\lambda)}$, $\tilde{F}^{(\chi)}
=F^{(\chi)}$ and $\tilde{F}^{(\eta)} =F^{(\eta)}$.

The existence of poles in the full vertex functions as $P\to 0$ is
also required by the Ward identities, discussed in
Sec.~\ref{Ward-id}. Moreover, the Ward identities alone allow us
to establish the explicit form of the poles. Indeed, by making use
of the relations for the amputated vertices in Eq.~(\ref{Ward-amp})
as well as the explicit form of the quark propagators in
Eqs.~(\ref{S_L}), (\ref{s_L}) and (\ref{s^C_L}), we obtain
\begin{mathletters}
\ba
\left. \Gamma^{A,i}_{aj,\mu}(k+P,k) \right|_{P\to 0}
&\simeq& \frac{P^{(\chi)}_{\mu}}{P^{\nu}P^{(\chi)}_{\nu}}
T^{A3}_{a} \delta^{i}_{j} \tilde{\Delta}_{k} {\cal P}_{+}, 
\label{pole-1}\\
\left. \Gamma^{A,ai}_{j,\mu}(k+P,k) \right|_{P\to 0}
&\simeq& -\frac{P^{(\lambda)}_{\mu}}{P^{\nu}P^{(\lambda)}_{\nu}}
T^{Aa}_{3} \delta^{i}_{j} \Delta_{k} {\cal P}_{-} , 
\label{pole-2}\\
\left. \Gamma^{A,i}_{j,\mu}(k+P,k) \right|_{P\to 0}
&\simeq& \frac{P^{(\eta)}_{\mu}}{P^{\nu}P^{(\eta)}_{\nu}}
\delta^{A}_{8} \delta^{i}_{j} \frac{\sqrt{3}}{2} 
\left[ \tilde{\Delta}_{k} {\cal P}_{+} 
-\Delta_{k} {\cal P}_{-} \right], \label{pole-3}\\
\left. \tilde{\Gamma}^{A,i}_{aj,\mu}(k+P,k) \right|_{P\to 0}
&\simeq& \frac{P^{(\chi)}_{\mu}}{P^{\nu}P^{(\chi)}_{\nu}}
T^{A3}_{a} \delta^{i}_{j} \tilde{\Delta}_{k} {\cal P}_{-}, 
\label{pole-4}\\
\left. \tilde{\Gamma}^{A,ai}_{j,\mu}(k+P,k) \right|_{P\to 0}
&\simeq& -\frac{P^{(\lambda)}_{\mu}}{P^{\nu}P^{(\lambda)}_{\nu}}
T^{Aa}_{3} \delta^{i}_{j} \Delta_{k}{\cal P}_{+} , 
\label{pole-5}\\
\left. \tilde{\Gamma}^{A,i}_{j,\mu}(k+P,k) \right|_{P\to 0}
&\simeq& \frac{P^{(\eta)}_{\mu}}{P^{\nu}P^{(\eta)}_{\nu}}
\delta^{A}_{8} \delta^{i}_{j} \frac{\sqrt{3}}{2} 
\left[ \tilde{\Delta}_{k} {\cal P}_{-} -
\Delta_{k} {\cal P}_{+} \right]. 
\label{pole-6}
\ea
\label{poles-LR}
\end{mathletters}
Now, by taking into account the definition of the amputated vertices
in Eq.~(\ref{amp-def}) and comparing the pole residues in
Eq.~(\ref{vertices-con}) with those in Eq.~(\ref{poles-LR}), we
unambiguously deduce the Dirac structure of the amputated (as well as
non-amputated) BS wave functions of the (pseudo-) NG bosons,
\begin{mathletters}
\ba 
\chi (p,0) &=& S^{-1}_{p} \bfchi (p,0) s^{-1}_{p} 
=\frac{\tilde{\Delta}_{p}}{F^{(\chi)}} {\cal P}_{+},
\label{chi-a} \\
\lambda (p,0) &=& s^{-1}_{p} \bflam (p,0) (p,0) S^{-1}_{p} 
=-\frac{\Delta_{p}}{F^{(\lambda)}} {\cal P}_{-}, 
\label{lambda-a}\\
\eta (p,0) &=& S^{-1}_{p} \bfeta (p,0) S^{-1}_{p} 
=\frac{\sqrt{3}}{F^{(\eta)}} \left( \tilde{\Delta}_{p} {\cal P}_{+}
-\Delta_{p} {\cal P}_{-} \right), \label{eta-a}\\
\tilde{\chi} (p,0) &=& \tilde{S}^{-1}_{p} \bftchi (p,0) \tilde{s}^{-1}_{p} 
=\frac{\tilde{\Delta}_{p}}{F^{(\chi)}}{\cal P}_{-}, 
\label{t-chi-a} \\
\tilde{\lambda} (p,0) &=& \tilde{s}^{-1}_{p} \bftlam (p,0) 
\tilde{S}^{-1}_{p} =-\frac{\Delta_{p}}{F^{(\lambda)}}{\cal P}_{+}, 
\label{t-lambda-a}\\
\tilde{\eta} (p,0) &=& \tilde{S}^{-1}_{p} \bfteta (p,0) \tilde{S}^{-1}_{p} 
=\frac{\sqrt{3}}{F^{(\eta)}} \left( \tilde{\Delta}_{p} {\cal P}_{-}
-\Delta_{p} {\cal P}_{+} \right). \label{t-eta-a} 
\ea
\end{mathletters}
This concludes our derivation. Before concluding this subsection, we
would like to emphasize that the arguments used here cannot be
generalized for the case of massive diquarks. The reason is that the
corresponding on-shell pole contributions to the vertex functions
[compare with Eq.~(\ref{vertices-con})] must appear at a
non-vanishing momentum $P$. Obviously, the structure of such poles
cannot be clarified by utilizing the Ward identities alone.

\subsection{NG doublet $\chi^{(\tilde{b})}_{a}$}
\label{doublet-chi}

Now, let us consider the BS equation for the massless $\chi$-doublet,
see Eq.~(\ref{chi}). As soon as the color symmetry is spontaneously 
broken in the model at hand, a non-trivial solution to this equation
should exist. In order to verify the self-consistency of our approach,
we have to check that this is the case. 

The most general Dirac structure of the amputated BS wave 
function, $\chi^{(\tilde{b})}_{a}(p,P) =S^{-1}_{p+P/2}
\bfchi^{(\tilde{b})}_{a}(p,P) s^{-1}_{p-P/2}$,
that is allowed by the space-time symmetries is given by
\ba
\chi_{a}^{(\tilde{b})} (p,P)&=&\delta_{a}^{~\tilde{b}}
\left[
 \chi_{1}^{-} \Lambda^{+}_{p}
+ \chi_{1}^{+} \Lambda^{-}_{p}
+ (p_{0}-\epsilon^{-}_{p}) \chi_{2}^{-} \gamma^0 \Lambda^{+}_{p}
+ (p_{0}+\epsilon^{+}_{p}) \chi_{2}^{+} \gamma^0 \Lambda^{-}_{p} 
\right. \nonumber\\
&+& \left. \chi_{3} (\vec{\gamma}\cdot\vec{P})
+ \chi_{4} (\vec{\alpha}\cdot\vec{P}) 
+ \chi_{5} \sigma^{nm}p^{n}P^{m} 
+ \chi_{6} \gamma^0\sigma^{nm}p^{n}P^{m} \right] {\cal P}_{+},
\label{chi-amp}
\ea
where $n,m=1,2,3$ are space indices, $\sigma^{nm}=i/2 [\gamma^{n},
\gamma^{m}]$, and the factors $(p_{0}-\epsilon^{-}_{p})$ and
$(p_{0}+\epsilon^{+}_{p})$ are introduced here for convenience. It is
of great advantage to notice that four out of eight independent
functions in this expression become irrelevant in the limit
$\vec{P}\to 0$. This agrees with the general statement made in
Subsection \ref{Deriv} (indeed, there are four independent Dirac
structures in the BS wave functions of spin zero states at zero
chemical potential \cite{FGMS}). We will consider only this limit
(which, in the case of NG bosons, implies that the total momentum
$P\to 0$).

After multiplying both sides of the BS equation (\ref{chi}) with
the appropriate quark propagators on the left and on the right, we
obtain the equation for the amputated BS wave function. This latter
splits into the following set of two equations:
\ba
{\cal P}_{+} \chi^{(\tilde{b})}_{a} (p,0) {\cal P}_{+} &=&
-\frac{2}{3} \pi \alpha_{s} \int
\frac{d^4 q}{(2\pi )^4} \gamma^{\mu}
{\cal P}_{-} \left( S_{q} \chi^{(\tilde{b})}_{a} (q,0) s_{q} 
+3 \bar{s}_{q} C \left(\chi^{(\tilde{b})}_{a} (-q,0)\right)^{T} 
C^{\dagger} S_{q} 
\right) {\cal P}_{-} \gamma^{\nu} {\cal D}_{\mu\nu}(q-p), \\
{\cal P}_{-} \chi^{(\tilde{b})}_{a} (p,0) {\cal P}_{+} &=&
\frac{2}{3} \pi \alpha_{s} \int
\frac{d^4 q}{(2\pi )^4} \gamma^{\mu}
{\cal P}_{+} S_{q} \chi^{(\tilde{b})}_{a} (q,0) s_{q} {\cal P}_{-}
\gamma^{\nu} {\cal D}_{\mu\nu}(q-p).
\ea
The quark propagators $S_{q}$ and $s_{q}$ are given in
Eqs.~(\ref{S_L}) and (\ref{s_L}). As we can see from those explicit
representations, not all of the terms are equally important. While
some of them develop large contributions in the vicinity of the Fermi
surface, the others are suppressed by powers of $\mu$. These latter
could be safely neglected in the leading order of the theory. Using
expression (\ref{D-long}) for the gluon propagator, we arrive at
the following approximate form of the BS equations:
\ba
{\cal P}_{+} \chi^{(\tilde{b})}_{a}(p,0) {\cal P}_{+} &\simeq&
\frac{2}{3} \pi \alpha_{s} \int
\frac{d^4 q}{(2\pi )^4} \gamma^{\mu}
{\cal P}_{-} \Lambda^{-}_{q} \left(
\frac{\gamma^0 (q_0-\epsilon_{q}^{-})-(\Delta^{-}_{q})^{*}}
{q_0^2-(\epsilon_{q}^{-})^2-|\Delta^{-}_{q}|^2}
\chi^{(\tilde{b})}_{a} (q,0) 
\frac{\gamma^0  }
{q_{0}-\epsilon_{q}^{-}} \right.
\nonumber \\
&&+\left.
\frac{3\gamma^0}
{q_{0}+\epsilon_{q}^{-}} 
\left(\chi^{(\tilde{b})}_{a} (-q,0)\right)^{T} 
\frac{\gamma^0 (q_0+\epsilon_{q}^{-})-(\Delta^{-}_{q})^{*}}
{q_0^2-(\epsilon_{q}^{-})^2-|\Delta^{-}_{q}|^2}
\right) \Lambda^{-}_{q} {\cal P}_{-}
\gamma^{\nu} {\cal D}_{\mu\nu}(q-p), \label{pp-pp-}\\
{\cal P}_{-} \chi^{(\tilde{b})}_{a}(p,0) {\cal P}_{+} &\simeq&
-\frac{2}{3} \pi \alpha_{s} \int
\frac{d^4 q}{(2\pi )^4} \gamma^{\mu}
{\cal P}_{+} \Lambda^{+}_{q}  
\frac{\gamma^0 (q_0+\epsilon_{q}^{-})-\Delta^{-}_{q}}
{q_0^2-(\epsilon_{q}^{-})^2-|\Delta^{-}_{q}|^2}
\chi^{(\tilde{b})}_{a} (q,0) 
\frac{\gamma^0 \Lambda^{-}_{q} {\cal P}_{-} }
{q_{0}-\epsilon_{q}^{-}}
\gamma^{\nu} {\cal D}_{\mu\nu}(q-p). \label{pp+pp+}
\ea
In the component form, these become
\ba
\chi_{1}^{-}(p) &\simeq& \frac{1}{3} \pi \alpha_{s} \int
\frac{d^4 q}{(2\pi )^4} \left[ \chi_{1}^{-}(q) 
+3 \chi_{1}^{-}(-q) - (\Delta^{-}_{q})^{*} \left(
\chi_{2}^{-}(q) +3\chi_{2}^{-}(-q)
\right)\right] \nonumber \\
&&\times
\frac{1}{q_0^2-(\epsilon_{q}^{-})^2-|\Delta^{-}_{q}|^2}
\mbox{tr}\left(\gamma^{\mu}
\Lambda^{-}_{q} \gamma^{\nu} \Lambda^{+}_{p} \right)
{\cal D}_{\mu\nu}(q-p), \label{chi-1}\\
(p_{0}-\epsilon^{-}_{p}) \chi_{2}^{-}(p) &\simeq& 
-\frac{1}{3} \pi \alpha_{s} \int \frac{d^4 q}{(2\pi )^4} 
\frac{\left[q_{0}^{2}-(\epsilon^{-}_{q})^2\right] \chi_{2}^{-}(q) 
- \Delta^{-}_{q} \chi_{1}^{-}(q) }{(q_{0}-\epsilon^{-}_{p})
\left[q_0^2-(\epsilon_{q}^{-})^2-|\Delta^{-}_{q}|^2\right]}
\nonumber \\
&&\times
\mbox{tr}\left(\gamma^{\mu} \Lambda^{+}_{q} 
\gamma^{0} \gamma^{\nu} \Lambda^{+}_{p} \gamma^{0}\right)
{\cal D}_{\mu\nu}(q-p) \label{chi-2},
\ea
plus the expressions that define $\chi^{+}_{i}$ in terms of
$\chi^{-}_{i}$. By noting that the equations for the even and odd
combinations of the BS wave functions, $\chi_{i}^{-}(p) \pm 
\chi_{i}^{-}(-p)$, decouple and satisfy the same kind of equation,  we
argue that it is sufficient for our purposes to consider only  the
even combination. Indeed, from the Ward identities, one finds that
$\chi^{-}_{1}(p)$ is related to the gap  $\Delta^{-}_{p}$. Then, if
$\chi^{-}_{1}(p)$ is odd, we must  have a non-trivial solution for the
gap satisfying $\Delta^{-}_{0}=\Delta^{-}_{p}|_{p=0}=0$. The analysis
of SD equation shows that no  such solution exists. Therefore, without
loss of generality,  we put $\chi_{i}^{-}(-p) =\chi_{i}^{-}(p)$, and
obtain
\ba
\chi_{1}^{-}(p) &\simeq& \frac{4}{3} \pi \alpha_{s} \int
\frac{d^4 q}{(2\pi )^4} \frac{ \chi_{1}^{-}(q) 
- (\Delta^{-}_{q})^{*} \chi_{2}^{-}(q) }
{q_0^2-(\epsilon_{q}^{-})^2-|\Delta^{-}_{q}|^2}
\mbox{tr}\left(\gamma^{\mu}
\Lambda^{-}_{q} \gamma^{\nu} \Lambda^{+}_{p} \right)
{\cal D}_{\mu\nu}(q-p), \label{chi-11}\\
(p_{0}-\epsilon^{-}_{p}) \chi_{2}^{-}(p) &\simeq& 
-\frac{1}{3} \pi \alpha_{s} \int \frac{d^4 q}{(2\pi )^4} 
\frac{\left[q_{0}^{2}-(\epsilon^{-}_{q})^2\right] \chi_{2}^{-}(q) 
- \Delta^{-}_{q}\chi_{1}^{-}(q) }{(q_{0}-\epsilon^{-}_{p})
\left[q_0^2-(\epsilon_{q}^{-})^2-|\Delta^{-}_{q}|^2\right]}
\mbox{tr}\left(\gamma^{\mu} \Lambda^{+}_{q} 
\gamma^{0} \gamma^{\nu} \Lambda^{+}_{p} \gamma^{0}\right)
{\cal D}_{\mu\nu}(q-p) \label{chi-22}.
\ea
By comparing our ansatz for the BS wave function in
Eq.~(\ref{chi-amp}) with the structure in Eq.~(\ref{chi-a}) that is
required by the Ward identities, we see that $\chi_{2}^{\pm}(p)$
components should be zero. The direct analysis of the BS equations, on
the other hand, shows that these component functions $\chi^{\pm}_{2}$ 
cannot be identically zero. It is not hard to pinpoint the origin of
the discrepancy. Indeed, in our approximation, we completely neglected
the wave function renormalization effects of quarks. Upon taking them
into account, the Ward identity (\ref{Ward-amp}) would lead to a
modified structure of the BS wave function, and all allowed Dirac
structures would be non-zero.

Therefore, as in the case of the wave function renormalization, we 
estimate the effect of $\chi_{2}^{-}(p)$ perturbatively. To this
end, we use $\chi_{2}^{-}(p)=0$ in the leading order of the theory.
Then, the equation for $\chi_{1}^{-}(p)$ reads
\ba
\chi_{1}^{-}(p) &\simeq& \frac{4}{3} \pi \alpha_{s} \int
\frac{d^4 q}{(2\pi )^4} \frac{ \chi_{1}^{-}(q) 
\mbox{ tr}\left(\gamma^{\mu}
\Lambda^{-}_{q} \gamma^{\nu} \Lambda^{+}_{p} \right)}
{q_0^2-(\epsilon_{q}^{-})^2-|\Delta^{-}_{q}|^2}
{\cal D}_{\mu\nu}(q-p). \label{chi-app}
\ea
On comparison with the gap equation (\ref{gap-eq}), we see that
$\chi_{1}^{-}(p) =(\Delta^{-}_{p})^{*}/F^{(\chi) }$ [as required by
the Ward identities, see Eq.~(\ref{chi-a})] is the exact solution to
the BS equation in the leading order approximation. Here, of course,
we assume that $\Delta^{-}_{p}$ is the solution to the gap equation.
By substituting the leading order solution $\chi_{1}^{-}(p)$ into
Eq.~(\ref{chi-22}), we get the estimate for $\chi_{2}^{-}(p) $.
In the most important region, $|\Delta^{-}_{0}|^{2} \alt p_{4}^{2}
+(\epsilon^{-}_{p})^{2}\alt \mu^{2}$, we find that 
\be 
\chi_{2}^{-}(p) \sim \frac{\alpha_{s} |\Delta^{-}_{0}|}
{F^{(\chi)} \sqrt{p_{4}^{2}+(\epsilon^{-}_{p})^{2}}}
\ln\frac{(2\mu)^{2}}{p_{4}^{2}+(\epsilon^{-}_{p})^{2}}.
\label{chi-2--}
\ee
Now, we also can check that this function could safely be neglected
in the equation for $\chi_{1}^{-}(p)$. Indeed, its substitution into
Eq.~(\ref{chi-11}) produces the result of order $\alpha_{s}
(\Delta^{-}_{p})^{*} /F^{(\chi)}$ which  is suppressed by a power of
$\alpha_{s}$ compared to $\chi_{1}^{-}(p) =(\Delta^{-}_{p})^{*}
/F^{(\chi)}$.

Therefore, both the corrections due to the wave function
renormalization of quarks \cite{Son,us,SW2,PR2,H1,Brown1,us2} and
those due to the non-vanishing component functions $\chi_{2}^{\pm}(p)$
are small in the leading order of the theory. Moreover, the
consistency with Ward identities requires that either both effects are
taken into account or neither of them.

\subsection{NG antidoublet $\lambda_{(\tilde{a})}^{b}$}
\label{doublet-lambda}

The analysis of the BS equation for the $\lambda$-antidoublet follows
very closely the analysis for the $\chi$-doublet. For the completeness
of the presentation, we still give all the details. 

The most general BS wave function of this antidoublet is given by
\ba
\lambda_{(\tilde{a})}^{b} (p,P)&=&\delta_{\tilde{a}}^{b}
{\cal P}_{-} \left[
\lambda_{1}^{+} \Lambda^{+}_{p}
+ \lambda_{1}^{-} \Lambda^{-}_{p}
+ (p_{0}-\epsilon^{-}_{p}) \lambda_{2}^{-} \gamma^0 \Lambda^{+}_{p}
+ (p_{0}+\epsilon^{+}_{p}) \lambda_{2}^{+} \gamma^0 \Lambda^{-}_{p} 
\right. \nonumber\\
&+& \left. \lambda_{3} (\vec{\gamma}\cdot\vec{P})
+ \lambda_{4} (\vec{\alpha}\cdot\vec{P}) 
+ \lambda_{5} \sigma^{nm}p^{n}P^{m} 
+ \lambda_{6} \gamma^0\sigma^{nm}p^{n}P^{m} \right] .
\label{lambda-amp}
\ea
As in the case of the $\chi$-doublet, to simplify the analysis we
restrict ourselves to the case of the vanishing total momentum $P\to
0$. Then, the equations for two projections of the wave function, 
${\cal P}_{-} \lambda_{(\tilde{a})}^{b}(p,0) {\cal P}_{-}$ and
${\cal P}_{-} \lambda_{(\tilde{a})}^{b}(p,0) {\cal P}_{+} $, 
read
\ba
{\cal P}_{-} \lambda_{(\tilde{a})}^{b} (p,0) {\cal P}_{-} &=&
-\frac{2}{3} \pi \alpha_{s} \int
\frac{d^4 q}{(2\pi )^4} \gamma^{\mu}
{\cal P}_{+} \left( s_{q} \lambda_{(\tilde{a})}^{b} (q,0) S_{q} 
+3 S_{q} C \left(\lambda_{(\tilde{a})}^{b} (-q,0)\right)^{T} 
C^{\dagger} \bar{s}_{q} 
\right) {\cal P}_{+} \gamma^{\nu} {\cal D}_{\mu\nu}(q-p), \\
{\cal P}_{-} \lambda_{(\tilde{a})}^{b} (p,0) {\cal P}_{+} &=&
\frac{2}{3} \pi \alpha_{s} \int
\frac{d^4 q}{(2\pi )^4} \gamma^{\mu}
{\cal P}_{+} s_{q} \lambda_{(\tilde{a})}^{b} (q,0) S_{q} {\cal P}_{-}
\gamma^{\nu} {\cal D}_{\mu\nu}(q-p).
\ea
After extracting the most significant in the vicinity of the Fermi
surface contributions from the quark propagators $S_{q}$ and $s_{q}$,
we arrive at
\ba
{\cal P}_{-} \lambda_{(\tilde{a})}^{b}(p,0) {\cal P}_{-} &\simeq&
\frac{2}{3} \pi \alpha_{s} \int
\frac{d^4 q}{(2\pi )^4} \gamma^{\mu}
{\cal P}_{+} \Lambda^{+}_{q} \left(
\frac{\gamma^0}
{q_{0}-\epsilon_{q}^{-}}
\lambda_{(\tilde{a})}^{b} (q,0) 
\frac{\gamma^0 (q_0-\epsilon_{q}^{-})-\Delta^{-}_{q}}
{q_0^2-(\epsilon_{q}^{-})^2-|\Delta^{-}_{q}|^2}
\right. 
\nonumber \\
&&+\left.
\frac{3\gamma^0 (q_0+\epsilon_{q}^{-})-\Delta^{-}_{q}}
{q_0^2-(\epsilon_{q}^{-})^2-|\Delta^{-}_{q}|^2}
\left(\lambda_{(\tilde{a})}^{b} (-q,0)\right)^{T} 
\frac{\gamma^0}
{q_{0}+\epsilon_{q}^{-}} 
\right) \Lambda^{+}_{q} {\cal P}_{+}
\gamma^{\nu} {\cal D}_{\mu\nu}(q-p), \label{ll-ll-}\\
{\cal P}_{-} \lambda_{(\tilde{a})}^{b}(p,0) {\cal P}_{+} &\simeq&
-\frac{2}{3} \pi \alpha_{s} \int
\frac{d^4 q}{(2\pi )^4} \gamma^{\mu}
\frac{{\cal P}_{+} \Lambda^{+}_{q}\gamma^0}
{q_{0}-\epsilon_{q}^{-}}
\lambda_{(\tilde{a})}^{b} (q,0) 
\frac{\gamma^0 (q_0+\epsilon_{q}^{-})-(\Delta^{-}_{q})^{*}}
{q_0^2-(\epsilon_{q}^{-})^2-|\Delta^{-}_{q}|^2}
{\cal P}_{-} \Lambda^{-}_{q}  
\gamma^{\nu} {\cal D}_{\mu\nu}(q-p). \label{ll+ll+}
\ea
Finally, rewriting this in components, we get 
\ba
\lambda_{1}^{-}(p) &\simeq& \frac{4}{3} \pi \alpha_{s} \int
\frac{d^4 q}{(2\pi )^4} \frac{ \lambda_{1}^{-}(q) 
- \Delta^{-}_{q} \lambda_{2}^{-}(q) }
{q_0^2-(\epsilon_{q}^{-})^2-|\Delta^{-}_{q}|^2}
\mbox{tr}\left(\gamma^{\mu}
\Lambda^{+}_{q} \gamma^{\nu} \Lambda^{-}_{p} \right)
{\cal D}_{\mu\nu}(q-p), \label{lambda-1}\\
(p_{0}-\epsilon^{-}_{p}) \lambda_{2}^{-}(p) &\simeq& 
-\frac{1}{3} \pi \alpha_{s} \int \frac{d^4 q}{(2\pi )^4} 
\frac{\left[q_{0}^{2}-(\epsilon^{-}_{q})^2\right] \lambda_{2}^{-}(q) 
- (\Delta^{-}_{q})^{*}\lambda_{1}^{-}(q) }{(q_{0}-\epsilon^{-}_{p})
\left[q_0^2-(\epsilon_{q}^{-})^2-|\Delta^{-}_{q}|^2\right]}
\mbox{tr}\left(\gamma^{\mu} \Lambda^{+}_{q} 
\gamma^{0} \gamma^{\nu} \Lambda^{+}_{p} \gamma^{0}\right)
{\cal D}_{\mu\nu}(q-p) \label{lambda-2},
\ea
where again, without loss of generality, we assumed that 
$\lambda^{-}_{i}(p)$ are even functions of momenta. 

By repeating the arguments of the previous subsection, we would find
that $\lambda_{2}^{-}(p) $ should be zero in a consistent
approximation when the wave function renormalizations of quarks are 
neglected. As in the case of the $\chi$-doublet, the equation for
$\lambda^{-}_{1}(p)$ component has the solution $\lambda^{-}_{1}(p)
=-\Delta^{-}_{p}/F^{(\lambda)}$ that is consistent  with the Ward
identities, see Eq.~(\ref{lambda-a}).

\subsection{NG singlets $\eta$}
\label{singlet}

The case of the massless singlets is very special. This is already
seen from the fact that the BS equations for the $\eta$ and
$\sigma$-singlets are coupled in general. This might appear
somewhat puzzling if one traces back the origin of the singlets.
While the $\eta$-singlet contains the antisymmetric tensor product
of two fundamental representations of $SU(3)$, the $\sigma$-singlet
comes from the product of the fundamental and the anti-fundamental
representations. Based on this observation, one might have concluded
that the bound state should form only in the $\eta$-channel. As we
shall see below, this argument is not completely groundless,
although, the real situation is slightly different. We would like to
point out that there is no symmetry in the color superconducting
phase of dense QCD which could prevent the coupling between the two
singlet channels. 

Let us consider the equation for the amputated BS wave functions of
singlets, $\eta(p,0)$ and $\sigma(p,0)$ (we consider only the case of
$P\to 0$ below). The most general structure of the wave functions is
\ba
\eta(p,0) &=& \left[
  \eta_{1}^{-} \Lambda^{+}_{p}
+ \eta_{1}^{+} \Lambda^{-}_{p}
+ (p_{0}-\epsilon^{-}_{p}) \eta_{2}^{-} \gamma^0 \Lambda^{+}_{p}
+ (p_{0}+\epsilon^{+}_{p}) \eta_{2}^{+} \gamma^0 \Lambda^{-}_{p}
\right] {\cal P}_{+} \nonumber\\
&&+\left[
  \eta_{3}^{+} \Lambda^{+}_{p}
+ \eta_{3}^{-} \Lambda^{-}_{p}
+ (p_{0}+\epsilon^{-}_{p}) \eta_{4}^{-} \gamma^0 \Lambda^{-}_{p}
+ (p_{0}-\epsilon^{+}_{p}) \eta_{4}^{+} \gamma^0 \Lambda^{+}_{p}
\right] {\cal P}_{-}, \label{eta-amp} \\
\sigma(p,0) &=&{\cal P}_{-} \gamma^0 \left[
(p_{0}-\epsilon^{-}_{p}) \sigma^{-} \Lambda^{+}_{p}
+(p_{0}+\epsilon^{+}_{p}) \sigma^{+} \Lambda^{-}_{p}
\right] {\cal P}_{+} . \label{sigma-amp}
\ea
Before proceeding any further, we would like to remind the definition  
of the $\eta$-singlet wave function in Eq.~(\ref{def-eta}). Unlike
other diquarks, it is built of only the Majorana spinors. By making
use of the generalized Majorana property in Eq.~(\ref{Maj-psi}), we
observe that the BS wave function $\eta(p,P)$ should satisfy the
following constraint:\footnote{Strictly speaking, from
Eq.~(\ref{def-eta})  one derives a relation for the non-amputated BS
wave function.  Assuming that the gap is an even function of
the momentum, it is  straightforward to show that the same relation
holds for the amputated wave function $\eta$.}
\be
C \eta^{T}(-p,P) C^{\dagger} = \eta(p,P) .
\ee
While rewritten in components, this restriction is satisfied when the
odd components $\eta_{1,3}^{\pm}(p)$ are even functions of momenta,
and when $\eta_{2}^{\pm}(-p)  = \eta_{4}^{\pm}(p)$.

The equations for different chiral projections of the amputated BS
wave functions, $\eta(p,0)=S^{-1}_{p}\bfeta(p,0)S^{-1}_{p}$ and 
$\sigma(p,0)=s^{-1}_{p}\bfsig(p,0)s^{-1}_{p}$, read 
\ba
{\cal P}_{+} \eta(p,0) {\cal P}_{+} &=&
-\frac{8}{3} \pi \alpha_{s} \int
\frac{d^4 q}{(2\pi )^4} \gamma^{\mu}
{\cal P}_{-} S_{q} \eta (q,0) S_{q} {\cal P}_{-}
\gamma^{\nu} {\cal D}_{\mu\nu}(q-p), \\
{\cal P}_{-} \eta(p,0) {\cal P}_{-} &=&
-\frac{8}{3} \pi \alpha_{s} \int
\frac{d^4 q}{(2\pi )^4} \gamma^{\mu}
{\cal P}_{+} S_{q} \eta (q,0) S_{q} {\cal P}_{+}
\gamma^{\nu} {\cal D}_{\mu\nu}(q-p), \\
{\cal P}_{-} \eta(p,0) {\cal P}_{+} &=&
-\frac{2}{3} \pi \alpha_{s} \int
\frac{d^4 q}{(2\pi )^4} \gamma^{\mu}
{\cal P}_{+} \left( 5 S_{q} \eta (q,0) S_{q} 
+6 s_{q} \sigma (q,0) s_{q} \right){\cal P}_{-}
\gamma^{\nu} {\cal D}_{\mu\nu}(q-p), \\
{\cal P}_{+} \eta(p,0) {\cal P}_{-} &=&
-\frac{2}{3} \pi \alpha_{s} \int
\frac{d^4 q}{(2\pi )^4} \gamma^{\mu}
{\cal P}_{-} \left( 5 S_{q} \eta (q,0) S_{q} 
+6 \bar{s}_{q} \left(\sigma (-q,0)\right)^{T} \bar{s}_{q}
\right) {\cal P}_{+}
\gamma^{\nu} {\cal D}_{\mu\nu}(q-p),\\
{\cal P}_{-} \sigma(p,0) {\cal P}_{+} &=&
-\frac{2}{3} \pi \alpha_{s} \int
\frac{d^4 q}{(2\pi )^4} \gamma^{\mu}
{\cal P}_{+} \left( 3 S_{q} \eta (q,0) S_{q} 
+2 s_{q} \sigma (q,0) s_{q} \right){\cal P}_{-}
\gamma^{\nu} {\cal D}_{\mu\nu}(q-p).
\ea
In components, these become
\ba
\eta_{1}^{-}(p) &\simeq& \frac{4}{3} \pi \alpha_{s} \int
\frac{d^4 q}{(2\pi )^4} \frac{ 
[q_{0}^{2}-(\epsilon^{-}_{q})^{2}]\eta_{1}^{-}(q) 
+ (\Delta_{q}^{-*})^{2} \eta_{3}^{-}(q) 
-2(\Delta^{-}_{q})^{*} [q_{0}^{2}-(\epsilon^{-}_{q})^{2}] 
\eta_{2}^{-}(q)}
{\left[q_0^2-(\epsilon_{q}^{-})^2-|\Delta^{-}_{q}|^2\right]^{2}}
\nonumber \\ &\times&
\mbox{tr}\left(\gamma^{\mu}
\Lambda^{-}_{q} \gamma^{\nu} \Lambda^{+}_{p} \right)
{\cal D}_{\mu\nu}(q-p), \label{eta-1}\\
\eta_{3}^{-}(p) &\simeq& \frac{4}{3} \pi \alpha_{s} \int
\frac{d^4 q}{(2\pi )^4} \frac{ 
[q_{0}^{2}-(\epsilon^{-}_{q})^{2}]\eta_{3}^{-}(q) 
+ (\Delta^{-}_{q})^{2} \eta_{1}^{-}(q) 
-2 \Delta^{-}_{q} [q_{0}^{2}-(\epsilon^{-}_{q})^{2}] \eta_{2}^{-}(q)}
{\left[q_0^2-(\epsilon_{q}^{-})^2-|\Delta^{-}_{q}|^2\right]^{2}}
\nonumber \\ &\times& \mbox{tr}\left(\gamma^{\mu}
\Lambda^{+}_{q} \gamma^{\nu} \Lambda^{-}_{p} \right)
{\cal D}_{\mu\nu}(q-p), \label{eta-3}\\
(p_{0}-\epsilon^{-}_{p}) \eta_{2}^{-}(p) &\simeq& 
\frac{1}{3} \pi \alpha_{s} \int \frac{d^4 q}{(2\pi )^4} 
\left( 5 \frac{
[q_{0}^{2}-(\epsilon^{-}_{q})^{2}] \eta_{2}^{-}(q)
+|\Delta^{-}_{q}|^2 \eta_{4}^{-}(q) -\Delta^{-}_{q} \eta_{1}^{-}(q)
-(\Delta^{-}_{q})^{*} \eta_{3}^{-}(q) }
{\left[q_0^2-(\epsilon_{q}^{-})^2-|\Delta^{-}_{q}|^2\right]^{2}}
\right. \nonumber \\ &+&\left. 
\frac{6 \sigma^{-}(q)}{q_{0}^{2}-(\epsilon^{-}_{q})^{2}}
\right) (q_{0}-\epsilon^{-}_{q}) 
\mbox{tr}\left(\gamma^{\mu} \Lambda^{+}_{q} 
\gamma^{0} \gamma^{\nu} \Lambda^{+}_{p} \gamma^{0}\right)
{\cal D}_{\mu\nu}(q-p) \label{eta-2}, \\
(p_{0}-\epsilon^{-}_{p}) \sigma^{-}(p) &\simeq& 
\frac{1}{3} \pi \alpha_{s} \int \frac{d^4 q}{(2\pi )^4} 
\left( 3 \frac{
[q_{0}^{2}-(\epsilon^{-}_{q})^{2}] \eta_{2}^{-}(q)
+|\Delta^{-}_{q}|^2 \eta_{4}^{-}(q) -\Delta^{-}_{q} \eta_{1}^{-}(q)
-(\Delta^{-}_{q})^{*} \eta_{3}^{-}(q) }
{\left[q_0^2-(\epsilon_{q}^{-})^2-|\Delta^{-}_{q}|^2\right]^{2}}
\right. \nonumber \\ &+&\left. 
\frac{2 \sigma^{-}(q)}{q_{0}^{2}-(\epsilon^{-}_{q})^{2}}
\right) (q_{0}-\epsilon^{-}_{q}) 
\mbox{tr}\left(\gamma^{\mu} \Lambda^{+}_{q} 
\gamma^{0} \gamma^{\nu} \Lambda^{+}_{p} \gamma^{0}\right)
{\cal D}_{\mu\nu}(q-p) \label{sigma-eq},
\ea
along with the irrelevant for our analysis expressions for the
plus-components. Note that we did not write down the equation for
$\eta_{4}^{-}(p)$ component since it is related to $\eta_{2}^{-}(p)$
as we argued above.

Now, let us analyze the BS equations for the singlets. By repeating
the argument involving the Ward identities, we see that the component
functions $\eta_{2,4}^{\pm}(p)$ should be {\em exactly} zero. As
opposed to the case of (anti-) doublet, a crucial difference appears
in the case of singlets. As we discussed in Subsec.~\ref{BS-str-NG},
the Dirac structure of the BS wave function of the $\eta$-singlet is
determined by the pole structure of the vertex in Eq.~(\ref{con-3}).
The explicit form of the latter is determined by the Ward identity in
Eq.~(\ref{ac}), and the result is presented in Eq.~(\ref{pole-3}).
The remarkable property of this result is that it does not get any
corrections even after the wave function renormalization effects of
quarks are taken into account. To see this, one should note that the
mentioned Ward identity (\ref{ac}) is given in terms of a single
propagator, $S_{p}$. Because of this, all wave function
renormalization effects always cancel from leading pole contribution
to the vertex of interest. 

It is very rewarding, therefore, to check that $\eta_{1}^{-}(p)=
-\sqrt{3} (\Delta^{-}_{p})^{*} /F^{(\eta)}$, $\eta_{3}^{-}(p)
=\sqrt{3} \Delta^{-}_{p} /F^{(\eta)}$ and $\eta_{2,4}^{-}(p)
=\sigma^{-}(p) =0$ is the exact solution to the BS
equations,\footnote{We believe that this is the only non-trivial
solution to the BS equation, although we were unable to rigorously
prove that no other solutions exist.} assuming that $\Delta_{p}$ is
the solution to the gap equation. It is interesting to notice that
no admixture of the $\sigma$-singlet appears in this solution for
the (pseudo-) NG boson.

\subsection{Decay constants}
\label{decay-consts}

In order to define the decay constants of $\chi_{a}^{(\tilde{b})}$,
$\lambda^{~b}_{(\tilde{a})}$, $\eta$ (pseudo-) NG bosons (as well as
their counterparts built of the right-handed fields), it is
convenient to introduce the following combinations of the currents:
\ba
j_{(a) \mu} &=& \sum_{A=4}^{7} T_{a}^{A3} j^{A}_{\mu},
\quad a=1,2,\\
j^{(b)}_{\mu} &=& \sum_{A=4}^{7} T_{3}^{Ab} j^{A}_{\mu}
\quad b=1,2 ,\\
j_{\mu} &=& j^{8}_{\mu},
\ea
where $j^{A}_{\mu}$ for $A=4,\ldots,8$ is defined in
Eq.~(\ref{current}). It is easy to check that the doublet
$\chi_{a}^{(\tilde{b})}$  couples only to $j_{(a) \mu}$, while the
antidoublet $\lambda^{~b}_{(\tilde{a})}$ couples only to
$j^{(b)}_{\mu}$. The singlet couples only to $j_{\mu}$. 

We will consider the low energy limit when the energy 
of the diquark (pseudo-) NG bosons $P_{0}\to 0$. Then, we define
their decay constants as follows: 
\ba
\langle 0| j_{(a) \mu} (0) | P, \tilde{b} \rangle_{L} 
&=& i \delta^{\tilde{b}}_{a} P^{(\chi)}_{\mu} F^{(\chi)}, \\
\langle 0| j^{(b)}_{\mu} (0) | P, \tilde{a} \rangle_{L} 
&=& i \delta_{\tilde{a}}^{b} P^{(\lambda)}_{\mu} F^{(\lambda)}, \\
\langle 0| j_{\mu}(0) | P \rangle_{L} 
&=& i P^{(\eta)}_{\mu} F^{(\eta)}, 
\ea
where $P^{(x)}_{\mu}=\left(P_{0}, -c_{x}^2 \vec{P}\right)$ and
$c_{x}$ is the velocity of (pseudo-) NG bosons. 

\begin{figure}
\epsfbox[25 230 400 370]{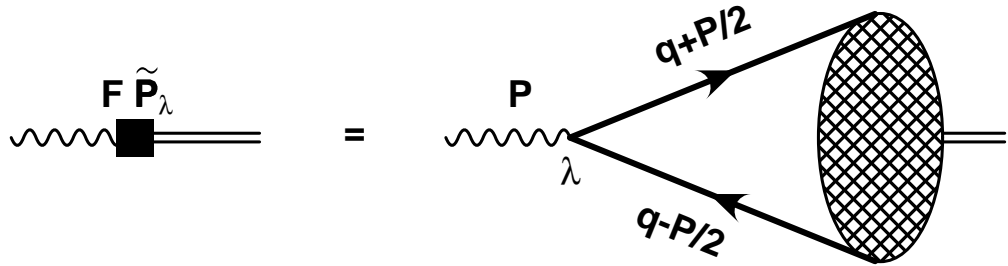}
\caption{The definition of the decay constant.}
\label{fig-decay}
\end{figure}

 From the definition of the current in Eq.~(\ref{current}) in terms 
of quark fields and from the definition of the BS wave functions in
Eq.~(\ref{BS-all-L}), we obtain 
\ba
P^{(\chi)}_{\mu} F^{(\chi)} &=& \frac{i}{2} \int 
\frac{d^{4}q}{(2\pi)^{4}}  \mbox{tr}
\left[\gamma_{\mu}{\cal P}_{+} \bfchi(q,P)\right]
 = \frac{i}{2} \int \frac{d^{4}q}{(2\pi)^{4}} \mbox{tr} \left[
\gamma_{\mu} {\cal P}_{+} S_{q+P/2} \chi(q,P) s_{q-P/2} \right], 
\label{decay-chi} \\ 
P^{(\lambda)}_{\mu} F^{(\lambda)} &=& \frac{i}{2} \int
\frac{d^{4}q}{(2\pi)^{4}} \mbox{tr}
\left[{\cal P}_{-} \gamma_{\mu} \bflam(q,P)\right]
 = \frac{i}{2} \int \frac{d^{4}q}{(2\pi)^{4}} \mbox{tr}\left[
{\cal P}_{-}  \gamma_{\mu} s_{q+P/2} \lambda(q,P) S_{q-P/2} \right],  
\label{decay-lambda}\\
P^{(\eta
)}_{\mu} F^{(\eta)} &=& \frac{i}{2\sqrt{3}} \int 
\frac{d^{4}q}{(2\pi)^{4}} \mbox{tr}
\left[\gamma_{\mu}{\cal P}_{+} \bfeta(q,P) \right]
= \frac{i}{2\sqrt{3}} \int 
\frac{d^{4}q}{(2\pi)^{4}} \mbox{tr} \left[\gamma_{\mu}
{\cal P}_{+} S_{q+P/2} \eta(q,P) S_{q-P/2} \right], 
\label{decay-eta}
\ea
where the doublet and the antidoublet BS wave functions are defined
so that $\bfchi_{a}^{(\tilde{b})}=\delta_{a}^{\tilde{b}}\bfchi $ 
and $\bflam^{~b}_{(\tilde{a})}=\delta^{~b}_{\tilde{a}}\bflam $. The
generic definition is diagrammatically presented in
Fig.~\ref{fig-decay}.


The definitions of the decay constants above are exact. The problem
is, however, that the solution for the BS wave functions $\chi(q,P)$,
$\lambda(q,P)$ and $\eta(q,P)$ at $P\neq 0$ is very hard to obtain. 
In order to get the estimates of the decay constants and velocities,
we will use the analogue of the Pagels-Stokar approximation \cite{PS}
(for a review see Ref.~\cite{Mir}). In this approximation, the wave
functions at  $P\neq 0$ are substituted by their values at $P=0$,
i.e., 
\ba
\chi(q,P) & \simeq & \chi(q,0) 
= \frac{\tilde{\Delta}_{q}}{F^{(\chi)}} {\cal P}_{+}, \\
\lambda(q,P) & \simeq & \lambda(q,0) 
= -\frac{\Delta_{q}}{F^{(\lambda)}} {\cal P}_{-} , \\
\eta(q,P)  & \simeq & \eta(q,0) 
= \frac{\sqrt{3}}{F^{(\eta)}} \left( \tilde{\Delta}_{q}{\cal P}_{+}
-\Delta_{q} {\cal P}_{-} \right),
\ea
where the right hand sides are fixed by the Ward identities.
By making use of this approximation and the explicit form of the 
quark propagators in Eqs.~(\ref{S_L}) and (\ref{s_L}), we 
derive in the weak coupling limit,
\ba
\left(F^{(\chi)}\right)^{2} 
\left\{ \begin{array}{c} P_{0} \\ c_{\chi}^{2} \vec{P} 
\end{array} \right\}
&\simeq & \frac{\mu^{2}}{16\pi^{3}}
\int_{0}^{1} x d x \int_{0}^{\pi} d \theta \sin\theta 
\left\{ \begin{array}{c} P_{0} \\ \vec{P} \cos^{2}\theta 
\end{array} \right\}
\int \frac{d q_{4} d \epsilon^{-}_{q} |\Delta^{-}_{q}|^{2}}
{\left[q_{4}^{2}+(\epsilon^{-}_{q})^{2}
+ x |\Delta^{-}_{q}|^{2}\right]^{2}}=
\frac{\mu^{2}}{8\pi^{2}}\left\{ \begin{array}{c} P_{0} \\ 
\frac{1}{3}\vec{P} \end{array} \right\}, \label{d-chi}\\
\left(F^{(\lambda)}\right)^{2} 
\left\{ \begin{array}{c} P_{0} \\ c_{\lambda}^{2} \vec{P} 
\end{array} \right\}
&\simeq & \frac{\mu^{2}}{16\pi^{3}}
\int_{0}^{1} x d x \int_{0}^{\pi} d \theta \sin\theta 
\left\{ \begin{array}{c} P_{0} \\ \vec{P} \cos^{2}\theta 
\end{array} \right\}
\int \frac{d q_{4} d \epsilon^{-}_{q} |\Delta^{-}_{q}|^{2}}
{\left[q_{4}^{2}+(\epsilon^{-}_{q})^{2} 
+ x |\Delta^{-}_{q}|^{2}\right]^{2}}=
\frac{\mu^{2}}{8\pi^{2}}\left\{ \begin{array}{c} P_{0} \\ 
\frac{1}{3}\vec{P} \end{array} \right\}, \label{d-lambda}\\
\left(F^{(\eta)}\right)^{2} 
\left\{ \begin{array}{c} P_{0} \\ c_{\eta}^{2} \vec{P} 
\end{array} \right\}
&\simeq & \frac{\mu^{2}}{8\pi^{3}}
\int_{0}^{\pi} d \theta \sin\theta 
\left\{ \begin{array}{c} P_{0} \\ \vec{P} \cos^{2}\theta 
\end{array} \right\}
\int \frac{d q_{4} d \epsilon^{-}_{q} |\Delta^{-}_{q}|^{2}}
{\left[q_{4}^{2}+(\epsilon^{-}_{q})^{2}+|\Delta^{-}_{q}|^{2}\right]^{2}}
=\frac{\mu^{2}}{8\pi^{2}}\left\{ \begin{array}{c} P_{0} \\ 
\frac{1}{3}\vec{P} \end{array} \right\}. \label{d-eta}
\ea
We observe that the decay constants of all NG bosons are of order
$\mu$, and all the velocities are equal to $1/\sqrt{3}$. This agrees
with similar results for the NG bosons in three flavor dense QCD
\cite{SonSt,CasGat,Rho,HZB}, as well as with the studies of 
Ref.~\cite{Risch}, dealing with the two flavor QCD in the framework
of the effective theory approach.
Notice that these
estimates of the decay constants imply that the infrared masses of the
five gluons are of order $g_{s}\mu$. (For a discussion of subtleties
concerning the generation of the mass of the eighth gluon, see
Ref.~\cite{CasDua}). 

For completeness of our presentation, let us mention that the 
above expressions for the decay constants in the Pagels-Stokar
approximation also contain the following subleading derivative
term corrections:
\ba
\delta\left[  
\left(F^{(\chi)}\right)^{2} \right]=
\delta\left[ \left(F^{(\lambda)}\right)^{2} \right]
&=& \frac{\mu^{2}}{4 \pi^{3}}
\int_{0}^{1} x(1-x) d x 
\int \frac{d q_{4} d \epsilon |\Delta^{-}_{q}|^{2}
q_{4}  \partial_{q_{4} } |\Delta^{-}_{q}|^{2} }
{\left[q_{4}^{2}+(\epsilon^{-}_{q})^{2}
+ x |\Delta^{-}_{q}|^{2}\right]^{3}} \nonumber \\
&-&\frac{\mu^{2}}{8 \pi^{3}}
\int_{0}^{1} d x 
\int \frac{d q_{4} d \epsilon q_{4} (3/4-x)
\partial_{q_{4}} |\Delta^{-}_{q}|^{2}}
{\left[q_{4}^{2}+(\epsilon^{-}_{q})^{2}
+ x |\Delta^{-}_{q}|^{2}\right]^{2}}, \label{cor-decay1}\\
\delta\left[ c_{\chi}^{2} \left(F^{(\chi)}\right)^{2} \right]
=\delta\left[ c_{\lambda}^{2} \left(F^{(\lambda)}\right)^{2} \right]
&=& \frac{\mu^{2}}{12 \pi^{3}}
\int_{0}^{1} x(1-x) d x 
\int \frac{d q_{4} d \epsilon |\Delta^{-}_{q}|^{2}
\epsilon  \partial_{\epsilon} |\Delta^{-}_{q}|^{2} }
{\left[q_{4}^{2}+(\epsilon^{-}_{q})^{2}
+ x |\Delta^{-}_{q}|^{2}\right]^{3}} \nonumber \\
&-&\frac{\mu^{2}}{24 \pi^{3}}
\int_{0}^{1} d x 
\int \frac{d q_{4} d \epsilon \epsilon (3/4-x)
\partial_{\epsilon} |\Delta^{-}_{q}|^{2}}
{\left[q_{4}^{2}+(\epsilon^{-}_{q})^{2}
+ x |\Delta^{-}_{q}|^{2}\right]^{2}},\label{cor-decay2} \\
\delta\left[ \left(F^{(\eta)}\right)^{2} \right]
&=& -\frac{\mu^{2}}{16 \pi^{3}}
\int \frac{d q_{4} d \epsilon 
q_{4}  \partial_{q_{4} } |\Delta^{-}_{q}|^{2} }
{\left[q_{4}^{2}+(\epsilon^{-}_{q})^{2}
+ |\Delta^{-}_{q}|^{2}\right]^{2}} , \\
\delta\left[ c_{\eta}^{2} \left(F^{(\eta)}\right)^{2} \right]
&=& -\frac{\mu^{2}}{48 \pi^{3}}
\int \frac{d q_{4} d \epsilon 
\epsilon \partial_{\epsilon} |\Delta^{-}_{q}|^{2} }
{\left[q_{4}^{2}+(\epsilon^{-}_{q})^{2}
+ |\Delta^{-}_{q}|^{2}\right]^{2}} ,
\ea
where $\epsilon\equiv \epsilon_{q} =|\vec{q}|-\mu$. In calculation, we
also assumed that the gap is an even function of $q_{4}$ and
$\epsilon$. By making use of the explicit solution to the gap equation
(see Sec.~\ref{SD-equation} and Appendix~B in Ref.~\cite{us}), we
could check that these corrections are suppressed by a power of the
coupling constant $\alpha_{s}$ compared to the leading order results
in Eqs.~(\ref{d-chi}), (\ref{d-lambda}) and (\ref{d-eta}). 

Here we worked out the decay constants of the left-handed diquark
composites. It is straightforward to repeat the analogous analysis
for the right-handed diquarks. As it should be, the results would be
the same, in agreement with the invariance under parity. 

It is worthwhile to note that the decay constants in the two
sectors of the theory could have been equivalently defined through
the matrix elements of the left- and right-handed color current. As
we discussed earlier, both such currents are approximately conserved
at large chemical potential. In other words, the latter means that
the corresponding vector and the axial vector currents are also
approximately conserved. Now, by making use of the definition of the
scalar and pseudoscalar diquarks in Eqs.~(\ref{scalar-def}) and
(\ref{pseudo-def}), it is easy to show that their decay constants
are equal, up to a factor of $\sqrt{2}$, to those of the left- and
right-handed states. Of course, here we assume that the definition
of the decay constants of scalars and pseudoscalars are given in
terms of the vector and axial vector currents, respectively.

While the NG scalars (a doublet, an antidoublet and a singlet) are
not the physical particles because of the Higgs mechanism, the five
pseudoscalars remain in the physical spectrum. Since the latter are
(nearly) massless in dense QCD, they should be relevant degrees of
freedom in the infrared dynamics. The decay constants of these
pseudoscalars are the physical observables that could be measured in
an experiment. The most likely decay products of these pseudoscalars
should be gluons of the unbroken $SU(2)_{c}$ and the massless quarks
of the third color (which might eventually get a small mass too if
another (non-scalar) condensate is generated \cite{W1,Sch-one}).

\section{BS equation for massive composites}
\label{BS-eq-mass}

The essential property of the quark pairing dynamics in $N_f=2$ dense
QCD is the long range interaction mediated by the gluons of the
magnetic type \cite{PR1,Son}. Of course, the Meissner effect in the
color superconducting phase produces masses for five out of the total
eight magnetic modes. Nevertheless, there are still three modes that
remain long ranged. This simple feature has many interesting
consequences. One of them was conjectured in Ref.~\cite{us3} where  it
was suggested that there should exist an infinite tower of massive
radial excitations in the diquark channels with the quantum numbers of
the (pseudo-) NG bosons. This conclusion was reached by making use of
an indirect argument based on special properties of the effective
potential in the color superconductor. 

In this section, we study the problem of massive radial excitations 
by using the rigorous approach of the BS equation, derived in
Sec.~\ref{Deriv-BS-eq}. As we shall see, the conjecture of
Ref.~\cite{us3} is essentially correct. At the same time, it will
turn out that some details of the pairing dynamics are rather
sensitive to the specific details (such as the Meissner effect) and
could not have been anticipated, based on the qualitative arguments
of Ref.~\cite{us3}.

\subsection{Bound states and the Meissner effect}
\label{bound-Meissner}

In this subsection, we would like to clarify the role of the Meissner
effect in the dynamics of diquark bound states. The main point we want
to emphasize is the existence of two classes of bound states, for
which the role of the Meissner effect is very different. The first
class consists of light bound states with the masses $M \ll
|\Delta_{0}^{-}|$. The binding energy of these states is large
(tightly bound states). The second class includes quasiclassical
states with the masses close to their threshold $k |\Delta_{0}^{-}|$
where $k=1$ for diquark doublets, built of one massive and one
massless quark, and $k=2$ for diquark singlets built of two massive
quarks. The binding energy of the quasiclassical states is small,
i.e., 
\be
\frac{k |\Delta_{0}^{-}|-M}{k |\Delta_{0}^{-}|} \ll 1.
\ee
It is clear that the quasiclassical bound states are sensitive to the
details of the infrared dynamics. Indeed, for long range potentials,
the quasiclassical part of the spectrum is almost completely
determined by the behavior of the potential at large distances. In the
particular case of cold dense QCD, the interaction  between quarks is
long ranged in the (imaginary) time direction  and essentially short
ranged in the spatial ones \cite{Son,us3}.  Because of that, the
region with $|k_{0}| < |\Delta^{-}_{0}| \alt  |\vec{k}|$ is
particularly important for the pairing dynamics  of the quasiclassical
diquark states. This implies that the  inclusion of the Meissner
effect is crucial for extracting  the properties of the states from
this second class (see  Appendix~\ref{diffequation}).

On the other hand, the Meissner effect is essentially irrelevant for
the light bound states. This point can be illustrated by the BS
equations for the lightest diquarks, the massless NG bosons. As was
shown in Sec.~\ref{BS-eq-NG}, the BS equations for them are
essentially equivalent to the gap equation.  And we know from the
experience of solving the gap equation 
\cite{us,SW2,PR2,H1,Brown1,us2} that the most important region of
momenta in the equation is given by $|\Delta_{0}^{-}| \alt |k_{0}| 
\ll |\vec{k}| \alt \mu$. In this particular region, the two kernels, 
with and without the Meissner effect taken into account [compare 
Eqs.~(\ref{D-long}) and  (\ref{Meissner})], are practically 
indistinguishable. Obviously, the same should be true for all very 
light bound states with $M \ll |\Delta_{0}^{-}|$.

In the rest of this section, we show that, because of the Meissner
effect, an infinite tower of (quasiclassical) massive diquarks
occurs only in the singlet channel. In the doublet channel, the only
bound states are the (tightly bound) NG bosons. This is connected
with the fact that the interaction in the doublet channel is
mediated exclusively by the gluons subject to the Meissner effect.
In the singlet channel, on the other hand, the interaction is partly
due to the unscreened gluons of the unbroken $SU(2)_{c}$ subgroup
and, therefore, the formation of massive (quasiclassical) states
is not prohibited.

\subsection{Massive doublet $\chi_{a}^{(\tilde{b})}$}
\label{chi-m-doublet}

Let us start from the analysis of the BS equation for the massive
$\chi$-doublets. By choosing the spatial component of the center of
mass momentum of the bound state zero, $P=(M_{\chi},0,0,0)$, we find
that, similarly to Eq.~(\ref{chi-amp}), the most general structure of
the BS wave function in the center of mass frame reads 
\ba
\chi_{a}^{(\tilde{b})} (p,M_{\chi})&=&\delta_{a}^{~\tilde{b}}
\left[
 \chi_{1}^{-} \Lambda^{+}_{p}
+ \chi_{1}^{+} \Lambda^{-}_{p}
+ (p_{0}-\epsilon^{-}_{p}+M_{\chi}/2) 
  \chi_{2}^{-} \gamma^0 \Lambda^{+}_{p}
+ (p_{0}+\epsilon^{+}_{p}+M_{\chi}/2) 
  \chi_{2}^{+} \gamma^0 \Lambda^{-}_{p} 
\right] {\cal P}_{+},
\label{chi-mass}
\ea
where, for convenience, we introduced here the factors 
$(p_{0}-\epsilon^{-}_{p}+M_{\chi}/2)$ and
$(p_{0}+\epsilon^{+}_{p}+M_{\chi}/2)$.
In contrast to the case of the (pseudo-) NG bosons, the structure of
the wave functions of massive states cannot be established from the
Ward identities. Although the vertex functions would also have the
poles, corresponding to the massive intermediate states, the Ward
identities are insufficient for extracting the structure of the
residues unambiguously. 

Be repeating the analysis similar to that in
Subsec.~\ref{doublet-chi}, we arrive at the following set of
equations for the components of the BS wave function:
\ba
\chi_{1}^{-}(p) &=&  \frac{4}{3} \pi \alpha_{s} \int
\frac{d^4 q}{(2\pi )^4}
\frac{ (q_0 -\epsilon^{-}_{q}+M_{\chi}/2) \left[\chi_{1}^{-}(q)
- (\Delta^{-}_{q})^{*} \chi_{2}^{-}(q) \right] }{
(q_0 -\epsilon^{-}_{q}-M_{\chi}/2) \left[
(q_0+M_{\chi}/2)^2 - (\epsilon^{-}_{q})^2 - |\Delta^{-}_{q}|^2\right]}
\nonumber\\ && \times \mbox{tr} \left[
\gamma^{\mu} \Lambda^{-}_{q} \gamma^{\nu}  \Lambda^{+}_{p}
\right] {\cal D}_{\mu\nu}(q-p), \label{chi1-m} \\
(p_0 -\epsilon^{-}_{p}+M_{\chi}/2)\chi_{2}^{-}(p) &=& 
- \frac{1}{3} \pi \alpha_{s} \int \frac{d^4 q}{(2\pi )^4}
\frac{ \left[(q_0 +M_{\chi}/2)^2 -(\epsilon^{-}_{q})^{2} \right] 
\chi_{2}^{-}(q) - \Delta^{-}_{q} \chi_{1}^{-}(q) }{
(q_0 -\epsilon^{-}_{q}-M_{\chi}/2) \left[ 
(q_0+M_{\chi}/2)^2 - (\epsilon^{-}_{q})^2 - |\Delta^{-}_{q}|^2\right]}
\nonumber\\
&& \times \mbox{tr} \left[ \gamma^{\mu} \gamma^0\Lambda^{-}_{q}
 \gamma^{\nu} \Lambda^{+}_{p} \gamma^0 \right]
{\cal D}_{\mu\nu}(q-p).
\label{chi2-m}
\ea
The other two equations express $\chi_{1,2}^{+}(p)$ in terms of
$\chi_{1.2}^{-}(p)$ and, therefore, they are irrelevant. As is easy
to check by tracing back the derivation of the BS equations, the
interaction in this doublet channel is provided exclusively by the
gluons that are subject to the Meissner effect in the color
superconducting phase. This fact could be taken into account
qualitatively by replacing the propagators of the magnetic modes
according to the qualitative rule in Eq.~(\ref{Meissner}). In
accordance with the discussion in Subsec.~\ref{bound-Meissner},
while such modification was irrelevant in the gap equation and in
the BS equations for the (pseudo-) NG bosons, it is going to play a
very important role in the analysis of the quasiclassical massive
states.

In order to proceed with the analysis of the BS equation, we will use
the approximation with $\chi_{2}^{\pm}=0$. Recall that such an 
approximation was completely justified in the case of the (pseudo-) NG
bosons. It is certain that it gives a very good approximation for
light bound states, with $M\ll |\Delta^{-}_{0}|$, in general. Our {\em
conjecture} is that the {\em ansatz} with $\chi_{2}^{\pm}=0$ yields a
reasonable approximation even in the case of quasiclassical states. In
order to justify this approximation one would need to prove that the
(perturbative) correction due to non-zero $\chi_{2}^{\pm}$ is small.
By repeating the arguments used for the (pseudo-) NG bosons, we could
show again that $\chi_{2}^{\pm}$ is irrelevant in the ultraviolet
region  $|\Delta^{-}_{0}| \alt p_{0} \alt \mu$. While this observation
is promising, it is not sufficient yet because the infrared region
$0 < p_{0} < |\Delta^{-}_{0}|$ is also important for the pairing
dynamics of the quasiclassical radial excitations. Therefore, for such
states, it should be considered as a (reasonable) conjecture.

Now, we drop $\chi_{2}^{\pm}$ from the analysis and assume that
the component functions $\chi_{1}^{\pm}$ depend only on the time
component of the momentum (compare with the analysis of the gap
equation in Refs.~\cite{Son,us,SW2,PR2,H1,Brown1,us2}). Then, we
arrive at the following equation: 
\begin{equation}
\chi_{1}^{-}(p_{4}) = \frac{2\alpha_{s}}{9\pi} \int_{0}^{\Lambda} 
d q_{4} K^{(\chi)}(q_{4}) \chi^{-}_{1} (q_{4}) 
\ln \frac{\Lambda}{|q_{4}-p_{4}|+c|\Delta^{-}_{0}|}, 
\label{BS-chi-mass}
\end{equation}
where $\Lambda=(4\pi)^{3/2}\mu/\alpha_{s}^{5/2}$ and the kernel
reads 
\begin{equation}
K^{(\chi)}(q_{4}) = \frac{ M_{\chi}^{2} 
\left( q_{4} +\sqrt{q_{4}^2 +|\Delta^{-}_{0}|^{2} } \right)^{2}
-|\Delta^{-}_{0}|^{4} } {\sqrt{q_{4}^2 +|\Delta^{-}_{0}|^{2} } 
\left[ 4 M_{\chi}^{2} q_{4}^2 
-\left( |\Delta^{-}_{0}|^{2} -M_{\chi}^{2} \right)^{2} \right]}.
\label{K-chi}
\end{equation}
To analyze the BS equation, we could convert it into a differential
equation, using the same approach as in the case of the gap equation
in Ref.~\cite{us} (see Appendix~\ref{diffequation}). It is
straightforward to show then that, in the ultraviolet region
$|\Delta^{-}_{0}| < p_{4} < \Lambda$, the BS wave function of the
massive doublet, $\chi_{1}^{-}(p_{4})$, has the same behavior as the
BS wave function of the NG boson in the $\chi$-channel (which, as we
know from the Ward identities, is proportional to the gap function).
The deviations might appear only in the infrared region, $0 < p_{4}
< |\Delta^{-}_{0}| $. Note, however, that the BS wave function in
that region is essentially constant. To see this, one should notice
that the dependence on $p_{4}$ on the right hand side of
Eq.~(\ref{BS-chi-mass}) comes through the argument of the
logarithm, in which either large $q_{4}$ or $c|\Delta^{-}_{0}|$
dominates.

By matching the infrared and the ultraviolet asymptotes of the
solution, we find that the non-trivial wave function
$\chi_{1}^{-}(p_{4})$ cannot have zeros in the whole range of momenta,
$0<p_{4}<\Lambda$. This means that the only solution to the integral
equation (\ref{BS-chi-mass}) corresponds to the NG  boson with
$M_{\chi}=0$. Indeed, a solution that describes a massive radial
excitation of the NG boson must have at least one zero. Because of the
Meissner effect, no such solutions exist. 

 From physical point of view, the massive gluons cannot provide a
sufficiently strong interaction to form massive radial excitations
of the NG bosons in the doublet channel. To understand this better,
it might be instructive to point out that, in absence of the
Meissner effect, an infinite tower of (quasiclassical) massive
states would appear (see Appendix~\ref{diffequation}). However, the
binding energy of all these states would be very small compared to
the value of the superconducting gap. This indicates that it is the 
long range dynamics that is primarily responsible for the formation
of such quasiclassical bound states. Then, in agreement with the
discussion in Subsec.~\ref{bound-Meissner}, even relatively small 
screening effects for gluons in the infrared region are sufficient
to prevent binding of quarks in the doublet channel.

\subsection{Massive antidoublet $\lambda_{(\tilde{a})}^{~b}$}
\label{lambda-m-doublet}

The analysis for the $\lambda_{(\tilde{a})}^{~b}$-antidoublet
resembles a lot the analysis in the previous subsection, so we 
outline only the main points, omitting the unnecessary details. 

The general structure of the BS wave function is given
\ba
\lambda_{(\tilde{a})}^{~b} (p,M_{\lambda})&=&\delta_{\tilde{a}}^{~b}
 {\cal P}_{-} \left[
 \lambda_{1}^{+} \Lambda^{+}_{p}
+ \lambda_{1}^{-} \Lambda^{-}_{p}
+ (p_{0}-\epsilon^{-}_{p}+M_{\lambda}/2) 
  \lambda_{2}^{-} \gamma^0 \Lambda^{+}_{p}
+ (p_{0}+\epsilon^{+}_{p}+M_{\lambda}/2) 
  \lambda_{2}^{+} \gamma^0 \Lambda^{-}_{p} 
\right].
\label{lambda-mass}
\ea
The equations for the components of the BS wave functions of the
antidoublet are almost the same as those for the doublet. The only
difference is that $M_{\chi}$ is replaced by $-M_{\lambda}$. By
repeating all the arguments of the previous subsection, we again
conclude that, after the Meissner effect is taken into consideration,
the gluon interaction is not strong enough to provide binding of the
massive radial excitations in the $\lambda_{(\tilde{a})}^{~b}$
antidoublet channel.

\subsection{Massive singlet $\eta$}
\label{eta-m-singlet}

Let us consider massive singlet diquark with $M_{\eta} \neq 0$. Since
the equation for $\eta$ and $\sigma$ do not decouple, the massive
radial excitation of the $\eta$ NG boson would have a non-zero
admixture of $\sigma$. The general structures of the BS wave functions
in the center of mass frame, $P=(M_{\eta},0,0,0)$, are 
\ba
\eta(p,M_{\eta}) &=& \left[
  \eta_{1}^{-} \Lambda^{+}_{p}
+ \eta_{1}^{+} \Lambda^{-}_{p}
+ (p_{0}-\epsilon^{-}_{p} +M_{\eta}/2) 
  \eta_{2}^{-} \gamma^0 \Lambda^{+}_{p}
+ (p_{0}+\epsilon^{+}_{p} +M_{\eta}/2)
  \eta_{2}^{+} \gamma^0 \Lambda^{-}_{p}
\right] {\cal P}_{+} \nonumber\\
&&+\left[
  \eta_{3}^{+} \Lambda^{+}_{p}
+ \eta_{3}^{-} \Lambda^{-}_{p}
+ (p_{0}+\epsilon^{-}_{p} -M_{\eta}/2)
  \eta_{4}^{-} \gamma^0 \Lambda^{-}_{p}
+ (p_{0}-\epsilon^{+}_{p} -M_{\eta}/2)
  \eta_{4}^{+} \gamma^0 \Lambda^{+}_{p}
\right] {\cal P}_{-}, \label{eta-amp-mass} \\
\sigma(p,M_{\eta}) &=& {\cal P}_{-} \gamma^0 \left[
  (p_{0}-\epsilon^{-}_{p} +M_{\eta}/2) 
  \sigma^{-} \Lambda^{+}_{p}
+(p_{0}+\epsilon^{+}_{p} +M_{\eta}/2)
  \sigma^{+} \Lambda^{-}_{p} \right] {\cal P}_{+} . 
\label{sigma-amp-mass}
\ea
The components satisfy the following set of equations:
\ba
 \eta_{1}^{-}(p) &=&  \frac{4}{3} \pi \alpha_{s} \int
\frac{d^4 q}{(2\pi )^4} \frac{1}{
\left[(q_0-M_{\eta}/2)^2 - (\epsilon^{-}_{q})^2 
      - |\Delta^{-}_{q}|^2\right]
\left[(q_0+M_{\eta}/2)^2 - (\epsilon^{-}_{q})^2 
      - |\Delta^{-}_{q}|^2\right]}
\nonumber\\
&\times& \Bigg\{\left[q_0^{2}
-(\epsilon^{-}_{q} -\frac{M_{\eta}}{2} )^{2}\right] 
\eta_{1}^{-}(q) +(\Delta_{q}^{-*})^{2} \eta_{3}^{-}(q) \nonumber \\
&&-(\Delta^{-}_{q})^{*} 
\left[q_0^{2} -(\epsilon^{-}_{q}-\frac{M_{\eta}}{2})^{2}
\right] \left(\eta_{2}^{-}(q) +\eta_{4}^{-}(q) \right)
\Bigg\} \mbox{tr} \left[
\gamma^{\mu} \Lambda^{-}_{q} \gamma^{\nu}  \Lambda^{+}_{p}
\right] {\cal D}_{\mu\nu}(q-p) , \label{eta1-mass} \\
\left(p_{0}+\frac{M_{\eta}}{2}-\epsilon^{-}_{p}\right)
\eta_{2}^{-}(p) &=& \frac{5}{3} \pi \alpha_{s} \int
\frac{d^4 q}{(2\pi )^4} \frac{1}{
\left[(q_0-M_{\eta}/2)^2 - (\epsilon^{-}_{q})^2 
      - |\Delta^{-}_{q}|^2\right]
\left[(q_0+M_{\eta}/2)^2 - (\epsilon^{-}_{q})^2 
      - |\Delta^{-}_{q}|^2\right]}
\nonumber\\
&\times& \Bigg\{
\left(q_0 +\frac{M_{\eta}}{2} +\epsilon^{-}_{q}\right) \left[\left(
q_0^{2} -(\epsilon^{-}_{q}-\frac{M_{\eta}}{2})^{2}\right) \eta_{2}^{-}(q)
-(\Delta^{-}_{q})^{*} \eta_{3}^{-}(q) \right] \nonumber\\
&&+\left(q_0 -\frac{M_{\eta}}{2} +\epsilon^{-}_{q}\right) 
\left(|\Delta^{-}_{q}|^2 \eta_{4}^{-}(q) 
-\Delta^{-}_{q} \eta_{1}^{-}(q) \right) \Bigg\}
\mbox{tr} \left[ \gamma^{0} \gamma^{\mu} \Lambda^{+}_{q} \gamma^{0} 
\gamma^{\nu} \Lambda^{+}_{p} \right] {\cal D}_{\mu\nu}(q-p)
 \nonumber \\
&+& 2 \pi \alpha_{s} \int
\frac{d^4 q}{(2\pi )^4} \frac{ \sigma^{-} }
{q_{0} -M_{\eta}/2-\epsilon^{-}_{q}}
\mbox{tr} \left[ \gamma^{0} \gamma^{\mu} \Lambda^{+}_{q} \gamma^{0} 
\gamma^{\nu} \Lambda^{+}_{p} \right] {\cal D}_{\mu\nu}(q-p), 
\label{eta2-mass} \\
 \eta_{3}^{-}(p) &=&   \frac{4}{3} \pi \alpha_{s} \int
\frac{d^4 q}{(2\pi )^4} \frac{1}{
\left[(q_0-M_{\eta}/2)^2 - (\epsilon^{-}_{q})^2 
      - |\Delta^{-}_{q}|^2\right]
\left[(q_0+M_{\eta}/2)^2 - (\epsilon^{-}_{q})^2 
      - |\Delta^{-}_{q}|^2\right]}
\nonumber\\
&\times& \Bigg\{
\left[q_0^{2} -(\epsilon^{-}_{q}+\frac{M_{\eta}}{2})^{2}\right] 
\eta_{3}^{-}(q) +(\Delta^{-}_{q})^{2} \eta_{1}^{-}(q) 
-\Delta^{-}_{q} 
\left[(q_0+\frac{M_{\eta}}{2})^{2} -(\epsilon^{-}_{q})^{2}\right] 
\eta_{2}^{-}(q) \nonumber \\
&&-\Delta^{-}_{q} 
\left[(q_0-\frac{M_{\eta}}{2})^{2} -(\epsilon^{-}_{q})^{2}\right] 
\eta_{4}^{-}(q) 
\Bigg\}\mbox{tr} \left[
\gamma^{\mu} \Lambda^{+}_{q} \gamma^{\nu}  \Lambda^{-}_{p}
\right] {\cal D}_{\mu\nu}(q-p), \label{eta3-mass}  \\
 \left(p_{0}-\frac{M_{\eta}}{2}+\epsilon^{-}_{p}\right)
\eta_{4}^{-}(p) &=&  \frac{5}{3} \pi \alpha_{s} \int
\frac{d^4 q}{(2\pi )^4} \frac{1}{
\left[(q_0-M_{\eta}/2)^2 - (\epsilon^{-}_{q})^2 
      - |\Delta^{-}_{q}|^2\right]
\left[(q_0+M_{\eta}/2)^2 - (\epsilon^{-}_{q})^2 
      - |\Delta^{-}_{q}|^2\right]}
\nonumber\\
&\times& \Bigg\{
\left(q_0 -\frac{M_{\eta}}{2} -\epsilon^{-}_{q}\right) \left[\left(
q_0^{2} -(\epsilon^{-}_{q}-\frac{M_{\eta}}{2})^{2}\right) \eta_{4}^{-}(q)
-(\Delta^{-}_{q})^{*} \eta_{3}^{-}(q) \right]\nonumber\\
&&+\left(q_0 +\frac{M_{\eta}}{2} -\epsilon^{-}_{q}\right) 
\left(|\Delta^{-}_{q}|^2 \eta_{2}^{-}(q) 
-\Delta^{-}_{q} \eta_{1}^{-}(q) \right) \Bigg\} 
\mbox{tr} \left[ \gamma^{0} \gamma^{\mu} \Lambda^{-}_{q} \gamma^{0} 
\gamma^{\nu} \Lambda^{-}_{p} \right] {\cal D}_{\mu\nu}(q-p)
\nonumber \\
&+& 2 \pi \alpha_{s} \int
\frac{d^4 q}{(2\pi )^4} \frac{ \sigma^{-}(-q) }
{q_{0} +M_{\eta}/2+\epsilon^{-}_{q}}
\mbox{tr} \left[ \gamma^{0} \gamma^{\mu} \Lambda^{-}_{q} \gamma^{0} 
\gamma^{\nu} \Lambda^{-}_{p} \right] {\cal D}_{\mu\nu}(q-p), 
\label{eta4-mass} \\
\left(p_{0}+\frac{M_{\eta}}{2}-\epsilon^{-}_{p}\right)
\sigma^{-}(p) &=& \pi \alpha_{s} \int
\frac{d^4 q}{(2\pi )^4} \frac{1}{
\left[(q_0-M_{\eta}/2)^2 - (\epsilon^{-}_{q})^2 
      - |\Delta^{-}_{q}|^2\right]
\left[(q_0+M_{\eta}/2)^2 - (\epsilon^{-}_{q})^2 
      - |\Delta^{-}_{q}|^2\right]}
\nonumber\\
&\times& \Bigg\{
\left(q_0 +\frac{M_{\eta}}{2} +\epsilon^{-}_{q}\right) \left[\left(
q_0^{2} -(\epsilon^{-}_{q}-\frac{M_{\eta}}{2})^{2}\right) \eta_{2}^{-}(q)
-(\Delta^{-}_{q})^{*} \eta_{3}^{-}(q) \right] \nonumber\\
&&+\left(q_0 -\frac{M_{\eta}}{2} +\epsilon^{-}_{q}\right) 
\left(|\Delta^{-}_{q}|^2 \eta_{4}^{-}(q) 
-\Delta^{-}_{q} \eta_{1}^{-}(q) \right) \Bigg\}
\mbox{tr} \left[ \gamma^{0} \gamma^{\mu} \Lambda^{+}_{q} \gamma^{0} 
\gamma^{\nu} \Lambda^{+}_{p} \right] {\cal D}_{\mu\nu}(q-p)
 \nonumber \\
&+& \frac{2}{3} \pi \alpha_{s} \int
\frac{d^4 q}{(2\pi )^4} \frac{ \sigma^{-} }
{q_{0} -M_{\eta}/2-\epsilon^{-}_{q}}
\mbox{tr} \left[ \gamma^{0} \gamma^{\mu} \Lambda^{+}_{q} \gamma^{0} 
\gamma^{\nu} \Lambda^{+}_{p} \right] {\cal D}_{\mu\nu}(q-p), 
\label{sigma-mass}
\ea
In the case of massless NG bosons, the component functions
$\eta^{-}_{2,4}$ equal zero. Similarly to the case of doublets, we
assume that the {\em ansatz} with these functions being equal to
zero yields a good approximation also for massive diquarks. 
By substituting $\eta^{-}_{2,4}=0$ into the BS equations above, we
obtain the following simple set of equations:
\ba
\eta_{1}^{-}(p) &=&  \frac{4}{3} \pi \alpha_{s} \int
\frac{d^4 q}{(2\pi )^4} \frac{ 
\left[q_0^{2} -(\epsilon^{-}_{q}-M_{\eta}/2)^{2}\right] 
\eta_{1}^{-}(q) 
+(\Delta_{q}^{-*})^{2} \eta_{3}^{-}(q) }{
\left[(q_0-M_{\eta}/2)^2 - (\epsilon^{-}_{q})^2 
      - |\Delta^{-}_{q}|^2\right]
\left[(q_0+M_{\eta}/2)^2 - (\epsilon^{-}_{q})^2 
      - |\Delta^{-}_{q}|^2\right]}
\nonumber \\ &&\times
\mbox{tr} \left[
\gamma^{\mu} \Lambda^{-}_{q} \gamma^{\nu}  \Lambda^{+}_{p}
\right] {\cal D}_{\mu\nu}(q-p), \label{eta1-m}  \\
\eta_{3}^{-}(p) &=&   \frac{4}{3} \pi \alpha_{s} \int
\frac{d^4 q}{(2\pi )^4} \frac{ 
\left[q_0^{2} -(\epsilon^{-}_{q}+M_{\eta}/2)^{2}\right] 
\eta_{3}^{-}(q) 
+(\Delta^{-}_{q})^{2} \eta_{1}^{-}(q) }{
\left[(q_0-M_{\eta}/2)^2 - (\epsilon^{-}_{q})^2 
      - |\Delta^{-}_{q}|^2\right]
\left[(q_0+M_{\eta}/2)^2 - (\epsilon^{-}_{q})^2 
      - |\Delta^{-}_{q}|^2\right]}
\nonumber \\ &&\times
\mbox{tr} \left[
\gamma^{\mu} \Lambda^{+}_{q} \gamma^{\nu}  \Lambda^{-}_{p}
\right] {\cal D}_{\mu\nu}(q-p), \label{eta3-m}  
\ea
plus the equation for $\sigma^{-}$ which does not allow a non-trivial
solution for a bound state. 

Now, in order to solve the set of equations for $\eta^{-}_{1}$ and
$\eta^{-}_{3}$, we make the following substitution:
\ba
\eta^{-}_{1}(p) = -\frac{(\Delta^{-}_{p})^{*}}{|\Delta^{-}_{p}|}
h_{1} (p) , \label{(3)} \\
\eta^{-}_{3}(p) = \frac{\Delta^{-}_{p}}{|\Delta^{-}_{p}|}
h_{3} (p) . \label{(4)}
\ea 
and, as in the case of doublets, we assume that the wave functions
($h_{1,3}$) depend only on the time component of the momentum
$p_{4}=ip_{0}$ (also compare with the analysis of the gap equation in
Refs.~\cite{Son,us,SW2,PR2,H1,Brown1,us2}). At the end, we arrive at
the following equation for the BS wave function ($h_{1}=h_{3}$) of
the massive singlet: 
\begin{equation}
h_{1}(p_{4}) = \frac{\alpha_{s}}{4\pi} \int_{0}^{\Lambda} 
d q_{4} K^{(\eta)}(q_{4}) h_{1} (q_{4}) 
\ln \frac{\Lambda}{|q_{4}-p_{4}|}, \label{BS-mass}
\end{equation}
where $\Lambda=(4\pi)^{3/2}\mu/\alpha_{s}^{5/2}$, and the kernel
reads 
\begin{equation}
K^{(\eta)}(q_{4}) = \frac{\sqrt{q_{4}^{2}+|\Delta^{-}_{0}|^{2}}}
{ q_{4}^{2}+|\Delta^{-}_{0}|^{2} -\left(M_{\eta}/2\right)^{2} }.
\label{K-eta}
\end{equation}
At this point it is appropriate to emphasize that, as we saw already
in the previous two subsections, the Meissner effect plays an
important role in the analysis of the massive bound states. Indeed,
our analysis indicates that only the long range interaction mediated
by the unscreened gluons of the unbroken $SU(2)_{c}$ is strong enough
to produce massive bound states. This is taken into account in
Eq.~(\ref{BS-mass}) where the effective coupling constant differs 
by the factor $9/8$ from the coupling in the gap equation
(\ref{sd-eq-appr}).

In order to get the solution for the BS wave function $h_{1}(p)$, we
use the same method as in the case of the gap equation \cite{us}. In
particular, we convert Eq.~(\ref{BS-mass}) into the differential
equation, 
\be 
p_{4} h_{1}^{\prime\prime}(p_{4})
+h_{1}^{\prime}(p_{4}) +\frac{\alpha_{s}}{4\pi} 
K^{(\eta)}(p_{4}) h_{1}(p_{4}) =0 , 
\label{dif-eq-eta}
\ee
along with the boundary conditions,
\be
h_{1}^{\prime}(0) =0    \quad \mbox{and} \quad 
h_{1}(\Lambda) = 0.
\label{bound-eta}
\ee
Now, we solve the differential equation (\ref{dif-eq-eta}) in the
following  three qualitatively different regions: $0 \leq p_{4} 
\leq \sqrt{|\Delta^{-}_{0}|^{2} -\left(M_{\eta}/2\right)^{2} }$,
$\sqrt{|\Delta^{-}_{0}|^{2} -\left(M_{\eta}/2\right)^{2} } \leq 
p_{4} \leq |\Delta^{-}_{0}|$ and $|\Delta^{-}_{0}| \leq p_{4} 
\leq \Lambda$. The kernel (\ref{K-eta}) has a simple behavior in 
each region, and the BS equation allows the analytical solutions, 
\begin{mathletters}
\ba
h_{1}(p_{4}) &=& 
C_{0} J_{0} \left(\sqrt{\frac{\alpha_{s} |\Delta^{-}_{0}| p_{4} }
{\pi \left[|\Delta^{-}_{0}|^{2}-\left(M_{\eta}/2\right)^{2}
\right]}}\right) ,  \quad \mbox{for} \quad 0 \leq p_{4} \leq 
\sqrt{|\Delta^{-}_{0}|^{2} -\left(M_{\eta}/2\right)^{2} },\\
h_{1}(p_{4}) &=& 
C_{1} J_{0} \left(\sqrt{\frac{\alpha_{s} |\Delta^{-}_{0}|}
{\pi p_{4}}}\right) 
+C_{2} N_{0} \left(\sqrt{\frac{\alpha_{s} |\Delta^{-}_{0}|} 
{\pi p_{4}}}\right) , \quad \mbox{for} \quad 
\sqrt{|\Delta^{-}_{0}|^{2} -\left(M_{\eta}/2\right)^{2} } 
\leq p_{4} \leq |\Delta^{-}_{0}| ,\\
h_{1}(p_{4}) &=& 
C_{3} \sin \left(\frac{1}{2} \sqrt{\frac{\alpha_{s}} {\pi}}
\ln\frac{\Lambda}{p_{4}}\right) , \quad \mbox{for} \quad 
|\Delta^{-}_{0}| \leq p_{4} \leq \Lambda ,
\ea
\end{mathletters}
where $J_{n}$ and $N_{n}$ are the Bessel functions of the first and
second type. The solutions are chosen so that the boundary
conditions are automatically satisfied. In the above expressions,
$C_{i}$ ($i=0,\ldots,3$) are the integration constants. To obtain the
spectrum of the massive diquark states, we match the logarithmic
derivatives of the appropriate pairs of the solutions at
$\sqrt{|\Delta^{-}_{0}|^{2} -\left(M_{\eta}/2\right)^{2} }$ and 
$|\Delta^{-}_{0}|$. After taking into account the equation that
determines the value of the gap (see Appendix~B in Ref.~\cite{us}), 
\be
\ln\frac{\Lambda}{|\Delta^{-}_{0}|} = \frac{2}{\nu}
\arctan\left(\frac{J_{0}(\nu)}{J_{1}(\nu)}\right), \quad
\nu=\sqrt{\frac{8\alpha_{s}}{9\pi}}, \label{eq-app-B}
\ee
the matching condition reads
\be
\frac{J_{0}(z_{0}) J_{1}(z_{0}) } 
{J_{1}(z_{0}) N_{0}(z_{0})+J_{0}(z_{0}) N_{1}(z_{0}) }
\simeq \frac{\sqrt{\alpha_{s}\pi}}{4}
\cot\left(\frac{3\pi}{4\sqrt{2}}\right), 
\label{condition}
\ee
where the coupling is assumed to be small, $\alpha_{s} \ll 1$, and
\be
z_{0} = \sqrt{\frac{\alpha_{s} |\Delta^{-}_{0}|}
{\pi\sqrt{|\Delta^{-}_{0}|^{2}-(M_{\eta}/2)^{2}}}}.
\ee
It is straightforward to check that the left hand side of
Eq.~(\ref{condition}) is an oscillating function having an infinite
number of zeros ($z_{0}\approx 2.40,~3.83,~5.52, \ldots$). In the
weakly coupled theory, each zero (or rather a nearby point that
approaches the zero as $\alpha_{s}\to 0$) determines a corresponding
value of the diquark mass\footnote{Note that there is also a zero at
$z_{0}=0$, but Eq.~(\ref{condition}) does not have a solution in its
vicinity.}. In the quasiclassical limit, i.e., when $M_{\eta} \to 2
|\Delta^{-}_{0}|$ from below, the left hand side of
Eq.~(\ref{condition}) is approximately given by $\cot(2z_{0})$. Then,
we derive the following simple estimates for the masses of the
$\eta$-singlets:
\begin{equation}
M^{2}_{n} \simeq 4 |\Delta^{-}_{0} |^{2} 
\left(1-\frac{2^{8}\alpha_{s}^{2}}{\pi^{6}(2n+1)^{4}}\right),
\quad n\gg 1.
\label{mass-sin-2}
\end{equation}
This agrees with the expression presented in Eq.~(\ref{mass-singlet})
when $\kappa = 2^{8}/\pi^{6}\approx 0.27$. Accidentally, one could
also check from the position of the zeros on the left hand side of
Eq.~(\ref{condition}) that the expression in Eq.~(\ref{mass-sin-2})
gives a good approximation even for the low lying states
($n=1,2,\ldots$). Notice that the state with $n=0$ does not appear.

\section{Conclusion}
\label{conclusion}

In this paper we studied the problem of diquark bound states in the
color superconducting phase of $N_f=2$ cold dense QCD. We used the
conventional method of BS equations that suits the problem best. We
derived the general BS equations, and then analyzed them in spin
zero channels. 

Our analytical analysis of the BS equations in cold dense QCD shows
that the theory contains five (nearly) massless pseudoscalars
(pseudo-NG bosons) which transform as a doublet, an antidoublet and
a singlet under the unbroken $SU(2)_{c}$. To the best of our
knowledge, these pseudoscalar diquarks have not been discussed in
the literature before. We estimate the decay constants of these
pseudoscalars, and find that their orders of magnitude are the same
as that of the chemical potential. The velocities of the
pseudoscalars are equal to $1/\sqrt{3}$, and this coincides with the
velocity of the NG bosons in three flavor QCD. While being (nearly)
massless, the five pseudoscalar diquarks should be the relevant
degrees of freedom in the low energy dynamics of $N_{f}=2$ dense QCD.

The parity-even partners of the pseudoscalar diquarks are the NG
bosons which are the ghosts in the theory. Although they are removed
from the spectrum of physical particles by the Higgs mechanism, one
cannot get rid of them completely, unless a special (unitary) gauge
is defined. Since the order parameter is given by a diquark
composite field, it does not seem to be straightforward to define
and to use the unitary gauge in dense QCD. In all the covariant
gauges we use here, the NG bosons are always present and they play
an important role in removing unphysical poles from physical
scattering amplitudes. 

We also studied the problem of massive diquarks. In accordance with
the conjecture of Ref.~\cite{us3}, there exists an infinite tower of
massive bound states which are the radial excitations of the
(pseudo-) NG bosons. As a result of the Meissner effect, it appears
that the massive radial excitations occur only in the singlet channel.
This could be understood in the following way. The interaction in the
doublet and the antidoublet channels is provided exclusively by the
gluons affected by the Meissner effect. Such interaction is not
sufficiently strong to form massive radial excitations in those
channels. The important point in this  analysis is the different role
the Meissner effect plays for tightly bound states and quasiclassical
bound states. 

As we know, the parity is unbroken in the color superconducting
phase of two flavor dense QCD. Then by  recalling that the left- and
right-handed sectors of the theory approximately decouple, we could
see that all the massive diquarks come in pairs of degenerate
parity-even (scalar) and parity-odd (pseudoscalar) states.

Regarding the nature of the massive diquark states, let us note that
they may truly be just resonances in the full theory, since they
could decay into the pseudo-NG bosons and/or gluons of the unbroken
$SU(2)_{c}$. At high density, however, both the running coupling 
$\alpha_{s}(\mu)$ and the effective Yukawa coupling $g_{Y} =
|\Delta^{-}_{0}|/F^{(x)} \sim |\Delta^{-}_{0}|/\mu$ are small, and,
therefore, these massive resonances are narrow.

At the end, we would like to add a few words about the higher spin
channels that we do not study here. In view of studies in
Ref.~\cite{PR2}, it would be of great interest to investigate also
the case of the spin one diquarks, as they might be rather light in
the color superconducting phase. The general form of the BS
equations for such diquarks are exactly the same as in
Eqs.~(\ref{chi}) through (\ref{sigma}). Of course, the structure of
the BS wave functions would differ.

\begin{acknowledgments}
V.A.M. is grateful to Koichi Yamawaki for his hospitality at Nagoya
University. The work of V.A.M. was partly supported by the 
Grant-in-Aid of Japan Society for the Promotion of Science 
No.~11695030. I.A.S. would like to thank D.~Rischke for valuable 
comments. The work of I.A.S. and L.C.R.W. was supported by the U.S. 
Department of Energy Grant No.~DE-FG02-84ER40153. 
\end{acknowledgments}

\appendix

\section{Angular integration}
\label{Ang-int}

We need to calculate the following traces over the Dirac indices:
\begin{eqnarray}
\mbox{tr}\left[\gamma^{\mu} \Lambda^{(e)}_{p}
\gamma^{\nu} \Lambda^{(e')}_{q}\right] &=&
g^{\mu\nu} (1 + e e' t )-2 e e' g^{\mu 0}g^{\nu 0} t
+ e e' \frac{\vec{q}^{\mu}\vec{p}^{\nu}+\vec{q}^{\nu}\vec{p}^{\mu}}
{|\vec{q}| |\vec{p}|}+\ldots, \label{tr1}\\
\mbox{tr}\left[\gamma^{\mu} \gamma^0 \Lambda^{(e)}_{p}
\gamma^{\nu} \gamma^0 \Lambda^{(e')}_{q}\right] &=&
-g^{\mu\nu} (1-e e' t )
+\left(g^{\mu 0} - e \frac{\vec{q}^{\mu}}{|\vec{q}|}\right)
\left(g^{\nu 0} - e' \frac{\vec{p}^{\nu}}{|\vec{p}|}\right)
+\left(g^{\mu 0} - e' \frac{\vec{p}^{\mu}}{|\vec{p}|}\right)
\left(g^{\nu 0} - e \frac{\vec{q}^{\nu}}{|\vec{q}|}\right),
\label{tr2}
\end{eqnarray}
where $e,e'=\pm 1$, $t =\cos\theta$ is the cosine of the angle between
three-vectors $\vec{q}$ and $\vec{p}$, and irrelevant antisymmetric
terms are denoted by the ellipsis.

By contracting these traces with the projectors of the magnetic,
electric and longitudinal types of gluon modes, we arrive at
\begin{eqnarray}
O^{(1)}_{\mu\nu} \mbox{~tr}\left[\gamma^{\mu} \Lambda^{(-)}_{p}
\gamma^{\nu}\Lambda^{(+)}_{q}\right] &=&2 (1-t )
\frac{q^2+p^2+q p(1-t )}
{q^2+p^2-2q pt }, \label{O01}\\
O^{(2)}_{\mu\nu} \mbox{~tr}\left[\gamma^{\mu}\Lambda^{(-)}_{p}
\gamma^{\nu}\Lambda^{(+)}_{q}\right] &=&2 (1+t )
\frac{q^2+p^2-q p(1+t )}{q^2+p^2-2q pt }
-(1+t ) \frac{(q-p)^2+(q_4-p_4)^2}{q^2+p^2-2q pt
+(q_4-p_4)^2}, \label{O02}\\
O^{(3)}_{\mu\nu} \mbox{~tr}\left[\gamma^{\mu}\Lambda^{(-)}_{p}
\gamma^{\nu}\Lambda^{(+)}_{q}\right] &=& (1+t )
\frac{(q-p)^2+(q_4-p_4)^2}
{q^2+p^2-2q pt +(q_4-p_4)^2}, \label{O03} \\
O^{(1)}_{\mu\nu} \mbox{~tr}\left[\gamma^{\mu} \Lambda^{(-)}_{p}
\gamma^{\nu}\Lambda^{(-)}_{q}\right] &=& 2 (1+t )
\frac{q^2+p^2-q p(1+t )}
{q^2+p^2-2q pt }, \label{O11}\\
O^{(2)}_{\mu\nu} \mbox{~tr}\left[\gamma^{\mu}\Lambda^{(-)}_{p}
\gamma^{\nu}\Lambda^{(-)}_{q}\right] &=&2 (1-t )
\frac{q^2+p^2+q p(1-t )}{q^2+p^2-2q pt }
-(1-t ) \frac{(q+p)^2+(q_4-p_4)^2}{q^2+p^2-2q pt
+(q_4-p_4)^2}, \label{O12}\\
O^{(3)}_{\mu\nu} \mbox{~tr}\left[\gamma^{\mu}\Lambda^{(-)}_{p}
\gamma^{\nu}\Lambda^{(-)}_{q}\right] &=& (1-t )
\frac{(q+p)^2+(q_4-p_4)^2}
{q^2+p^2-2q pt +(q_4-p_4)^2}, \label{O13} \\
O^{(1)}_{\mu\nu} \mbox{~tr}\left[
\gamma^{\mu} \gamma^0 \Lambda^{(-)}_{p}
\gamma^{\nu} \gamma^0 \Lambda^{(-)}_{q}\right] &=&
-2 (1-t ) \frac{q^2+p^2+q p(1-t )}
{q^2+p^2-2q pt }, \label{O21}\\
O^{(2)}_{\mu\nu} \mbox{~tr}\left[
\gamma^{\mu} \gamma^0 \Lambda^{(-)}_{p}
\gamma^{\nu} \gamma^0 \Lambda^{(-)}_{q}\right] &=&
\frac{2q p(1-t^2 )}{q^2+p^2-2q pt } + (1+t )
\frac{(q-p)^2 -(q_4-p_4)^2 -2i(q_4-p_4)(q-p)}
{q^2+p^2-2q pt +(q_4-p_4)^2}, \label{O22}\\
O^{(3)}_{\mu\nu} \mbox{~tr}\left[
\gamma^{\mu} \gamma^0 \Lambda^{(-)}_{p}
\gamma^{\nu} \gamma^0 \Lambda^{(-)}_{q}\right] &=& -(1+t )
\frac{(q-p)^2 -(q_4-p_4)^2 -2i(q_4-p_4)(q-p)}
{q^2+p^2-2q pt +(q_4-p_4)^2}, \label{O23} \\
O^{(1)}_{\mu\nu} \mbox{~tr}\left[
\gamma^{\mu} \gamma^0 \Lambda^{(-)}_{p}
\gamma^{\nu} \gamma^0 \Lambda^{(+)}_{q}\right] &=&
-2 (1+t ) \frac{q^2+p^2-q p(1+t )}
{q^2+p^2-2q pt }, \label{O31}\\
O^{(2)}_{\mu\nu} \mbox{~tr}\left[
\gamma^{\mu} \gamma^0 \Lambda^{(-)}_{p}
\gamma^{\nu} \gamma^0 \Lambda^{(+)}_{q}\right] &=&
-\frac{2q p(1-t^2 )}{q^2+p^2-2q pt } + (1-t )
\frac{(q+p)^2 -(q_4-p_4)^2 -2i(q_4-p_4)(q+p)}
{q^2+p^2-2q pt +(q_4-p_4)^2}, \label{O32}\\
O^{(3)}_{\mu\nu} \mbox{~tr}\left[
\gamma^{\mu} \gamma^0 \Lambda^{(-)}_{p}
\gamma^{\nu} \gamma^0 \Lambda^{(+)}_{q}\right] &=& -(1-t )
\frac{(q+p)^2 -(q_4-p_4)^2 -2i(q_4-p_4)(q+p)}
{q^2+p^2-2q pt +(q_4-p_4)^2}, \label{O33}
\end{eqnarray}
where $q\equiv |\vec{q}|$, $p\equiv |\vec{p}|$, $q_4\equiv -i q_0$
and $p_4\equiv -i p_0$.

Therefore,
\begin{eqnarray}
I_{1}^{-+}&=&q^2 \int d\Omega
{\cal D}_{\mu\nu}(q-p) \mbox{~tr}\left[ \gamma^{\mu}
\Lambda^{(-)}_{p} \gamma^{\nu} \Lambda^{(+)}_{q}
\right] \nonumber \\
&&\approx 2i\pi \left[
\frac{2}{3}\ln\frac{(2\mu)^3}{|\epsilon_{q}^{-}|^3+\pi M^2 \omega/2}
+\ln\frac{(2\mu)^2}{(\epsilon_{q}^{-})^2+2M^2+\omega^2}
+\xi\right] , \label{angle11}\\
I_{1}^{--}&=&q^2 \int d\Omega
{\cal D}_{\mu\nu}(q-p) \mbox{~tr}\left[ \gamma^{\mu}
\Lambda^{(-)}_{p} \gamma^{\nu} \Lambda^{(-)}_{q}
\right] \nonumber \\
&&\approx 2i\pi \left[
\frac{2}{3}\ln\frac{(2\mu)^3}{|\epsilon_{q}^{-}|^3+\pi M^2 \omega/2}
-\frac{\alpha_{s}}{\pi}\ln\frac{(2\mu)^2}
{(\epsilon_{q}^{-})^2+2M^2+\omega^2}
+\xi \ln\frac{(2\mu)^2}{(\epsilon_{q}^{-})^2+\omega^2}\right] , 
\label{angle12}
\end{eqnarray}
where $M^2=2\alpha_{s}\mu^2/\pi$ and $\omega=|q_4-p_4|$.
\begin{eqnarray}
I_{2}^{--}&=&q^2 \int d\Omega
{\cal D}_{\mu\nu}(q-p) \mbox{~tr}\left[\gamma^{\mu} \gamma^0
\Lambda^{(-)}_{p} \gamma^{\nu} \gamma^0 \Lambda^{(-)}_{q}
\right] \nonumber \\
&& \approx 2i\pi \left[
-\frac{2}{3}\ln\frac{(2\mu)^3}{|\epsilon_{q}^{-}|^3+\pi M^2 \omega/2}
+\ln\frac{(2\mu)^2}{(\epsilon_{q}^{-})^2+2M^2+\omega^2}
- \xi \right] ,
\label{angle21} \\
I_{2}^{-+}&=&q^2 \int d\Omega
{\cal D}_{\mu\nu}(q-p) \mbox{~tr}\left[\gamma^{\mu} \gamma^0
\Lambda^{(-)}_{p} \gamma^{\nu} \gamma^0 \Lambda^{(+)}_{q}
\right] \nonumber \\
&& \approx 2i\pi \left[
-\frac{2}{3}\ln\frac{(2\mu)^3}{|\epsilon_{q}^{-}|^3+\pi M^2 \omega/2}
+\ln\frac{(2\mu)^2}{(\epsilon_{q}^{-})^2+2M^2+\omega^2}
- \xi \ln\frac{(2\mu)^2}{(\epsilon_{q}^{-})^2+\omega^2}\right] ,
\label{angle22}
\end{eqnarray}

\section{A non-perturbative correction to the SD equation}
\label{non-pert}

\begin{figure}
\epsfbox[25 220 400 380]{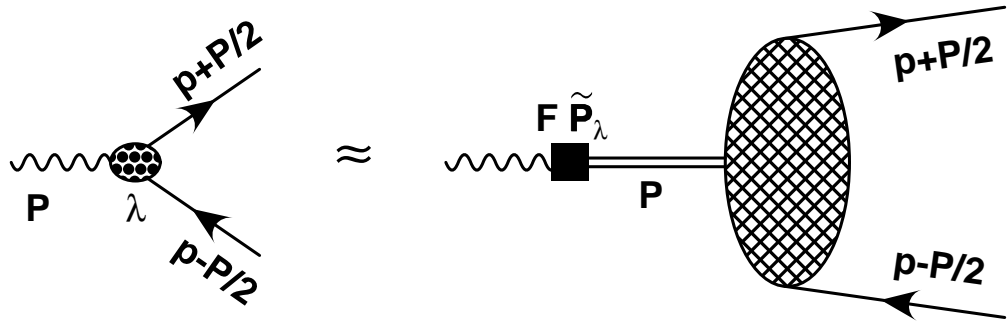}
\caption{The pole contribution to the vertex as $P \to 0$.}
\label{fig-ver}
\end{figure}

In light of our analysis in Sec.~\ref{Ward-id}, one could argue that
the SD equation might get a large non-perturbative contribution,
coming from the pole contributions in the full vertex function, see 
Eq.~(\ref{poles-LR}) and Fig.~\ref{fig-ver}. We remind that the pole
structure of the vertices is related to the  existence of the NG and
pseudo-NG bosons in the theory (for more on this see
Sec.~\ref{BS-eq-NG}). 


If one recalls that the SD equation is quite sensitive to the long
range dynamics ($|P| \ll \mu$), it would be very natural to ask
whether the pole contributions to the vertex function in 
Eqs.~(\ref{poles-LR}) could modify the SD equation and its solution.
The revealed non-perturbative contributions could conveniently be 
combined in the matrix form as follows:
\be
\left. \delta \Gamma^{A\mu}(q+P,q) \right|_{P\to 0}
=\frac{\tilde{P}^{\mu}}{P^{\nu}\tilde{P}_{\nu}} 
\left(\begin{array}{ccc}
3 \delta^{i}_{~j} \delta^{A}_{8} (T^{8})_{a}^{~b} 
\left(\Delta_{q}{\cal P}_{-}-\tilde{\Delta}_{q}{\cal P}_{+}\right) &
\delta^{i}_{~j} (T^{A})_{a}^{~3} \tilde{\Delta}_{q} {\cal P}_{+}  &
-\hat{\varepsilon}_{ac}^{ij} (T^{A})_{3}^{~c} \Delta_{q} {\cal P}_{-} \\
- \delta^{i}_{~j} (T^{A})_{3}^{~b} \Delta_{q} {\cal P}_{-}&
0 & 0 \\
\hat{\varepsilon}^{cb}_{ij} (T^{A})_{c}^{~3} 
\tilde{\Delta}_{q} {\cal P}_{+}  & 0 & 0
\end{array}\right). \label{vert-cor}
\ee
It is of great importance to notice that this is longitudinal, i.e.,
$\delta \Gamma^{A\mu} (q+P,q) \sim \tilde{P}^{\mu} \equiv
(P^{0},\vec{P}/3)$. As  a result of this property, the contraction of
this vertex with the transverse (with respect to $\vec{P}$) projector
of the magnetic gluon modes is equal to zero. Regarding the other two
types of the gluon modes (electric and longitudinal), the
corresponding contractions are non-zero, and they lead to a finite
contribution to the right hand side of the SD equation (\ref{SD-eq}).
We stress that the vertex in Eq.~(\ref{vert-cor}) is longitudinal
with respect to $\tilde{P}^{\mu}$ (notice tilde), while the projector of
the electric modes is transverse with respect to $P^{\mu}$ (no tilde
here). This difference is responsible for a non-zero contraction
involving the electric gluon modes. It is still, however, the
longitudinal gluon mode that, after being contracted with the vertex
in Eq.~(\ref{vert-cor}), gives the most significant contribution to
the SD equation. By performing the explicit calculation, we arrive at
the following extra term to the right hand side of the gap equation:
\be
\delta \Delta^{-}_{p} \simeq  
-\frac{\pi}{2} \alpha_{s}\int\frac{d^4 q}{(2\pi)^4}
\mbox{tr} \left( \gamma^{\mu} \Lambda^{+}_{q} \gamma^{0}
\Lambda^{-}_{p} \right)
\frac{(q_0+\epsilon_{q}^{-}) \Delta^{-}_{q}}
{q_0^2-(\epsilon_{q}^{-})^2-|\Delta^{-}_{0}|^2}
\frac{(\tilde{q}-\tilde{p})^{\nu}}
{(\tilde{q}-\tilde{p})^{\lambda}(q-p)_{{\lambda}}} 
{\cal D}_{\mu\nu}(q-p),
\label{gap-cor}
\ee
which results in the following term to the equations for 
$\Delta^{-}_{p}$:
\ba
\delta \Delta^{-}_{p} &\simeq &
\frac{\alpha_{s}}{16 \pi^2} \int d q_{4} d q
\frac{\Delta^{-}_{q}}
{q_4^2+(\epsilon_{q}^{-})^2+|\Delta^{-}_{0}|^2}
\left[-\xi + O\left(\frac{q_4^2+(\epsilon_{q}^{-})^2}{M^2}
\ln\frac{(2\mu)^2}{q_4^2+(\epsilon_{q}^{-})^2} \right) 
\right], \nonumber \\
&\simeq & \frac{\alpha_{s}}{16 \pi} \int 
\frac{d q_{4} \Delta^{-}_{q}}
{\sqrt{q_4^2+|\Delta^{-}_{0}|^2}}
\left[-\xi + O\left(\frac{q_4^2}{M^2}\right) 
\right],
\label{del-cor}
\ea
This correction is of the same order as the correction from the
longitudinal gluon modes in the gap equation (\ref{gap-eq}) when the
bare vertices are used \cite{us}. Therefore, such an additional
correction could only modify the overall constant factor in the
solution for $\Delta^{-}_{p}$. The exponential factor and the
overall power of the  coupling constant in the solution should
remain intact.

Of course, as we discussed earlier, the overall constant in the
expression for the gap might get other kinds of corrections which
have not been analyzed here \cite{Brown1,0004013}. Sorting out all
such corrections is a rather complicated problem that is outside the
scope of this paper.

\section{The analytical solutions to the BS equations}
\label{diffequation}

In this appendix, we present the approximate analytical solutions
to the BS equations. In general, our approach here resembles 
the method commonly used for solving the gap equation
\cite{us,SW2,PR2,H1,Brown1,us2}. One of the purposes of the analysis
below is to illustrate that, while the Meissner effect is irrelevant
for the pairing dynamics of light bound states with $M \ll
|\Delta_{0}^{-}|$, it is crucial for the pairing of quasiclassical
bound states (see Subsec.~\ref{bound-Meissner}).

\subsection{NG bosons}
\label{diff-eq-NG}

Let us start from the BS equation (\ref{chi-app}) for the (pseudo-)
NG bosons in the $\chi$-doublet channel. After performing the
standard approximations that were extensively discussed in many 
studies of the gap equation \cite{us,SW2,PR2,H1,Brown1,us2} (see
also Sec.~\ref{SD-equation}), we arrive at the following integral
equation:
\begin{equation}
\chi(p_{4})\simeq\frac{2\alpha_{s}}{9\pi}
\int_{0}^{\Lambda}  \frac{d q_{4} \chi (q_{4})}
{\sqrt{q_{4}^2+|\Delta^{-}_{0}|^2}}
\ln\frac{\Lambda}{|p_{4}-q_{4}|+c |\Delta_{0}^{-}| },
\label{chi-c1}
\end{equation}
where, for brevity of notation, we use $\chi\equiv \chi_{1}^{-}$ and
$\Lambda = (4\pi)^{3/2} \mu/\alpha_{s}^{5/2}$. Notice that, in
accordance with Eq.~(\ref{Meissner}), the Meissner effect is taken
into account by the term $c |\Delta_{0}^{-}|$ in the logarithm.
Without loss of generality, we substitute $c=1$ in what
follows\footnote{The analysis for $c\neq 1$ is a little more
complicated since the two scales, $c|\Delta_{0}^{-}|$ and
$|\Delta_{0}^{-}|$, are different. Despite this, the final result
for $c\neq 1$ would remain qualitatively the same as soon as $c$ is
a constant of order one.}. The integral equation (\ref{chi-c1})
could be approximately reduced to the following second order
differential equation:
\be 
(p_{4}+|\Delta_{0}^{-}|) \chi^{\prime\prime}(p_{4})
+\chi^{\prime}(p_{4}) +\frac{\nu^{2}}{4}
\frac{\chi(p_{4})}{\sqrt{p_{4}^{2}+|\Delta_{0}^{-}|^{2}}} =0 , 
\quad \mbox{where} \quad  \nu=\sqrt{\frac{8\alpha_{s}}{9\pi}}, 
\label{chi-c2}
\ee
subject to the infrared and the ultraviolet boundary conditions,
\be
\chi^{\prime}(0) =0    \quad \mbox{and} \quad 
\chi(\Lambda) -|\Delta_{0}^{-}| \chi^{\prime}(\Lambda)
\approx 0.
\label{bound-c}
\ee
In order to get the estimate for the solution, we consider the
differential equation on the two adjacent intervals, $0 \leq p_{4}
\leq |\Delta_{0}^{-}|$ and $|\Delta_{0}^{-}| \leq p_{4} \leq
\Lambda$, separately. The approximate analytical solutions that
satisfy the boundary conditions in Eq.~(\ref{bound-c}) read
\begin{mathletters}
\ba
\chi(p_{4}) &=& C_{1} 
\exp\left(\frac{p_{4}}{2|\Delta_{0}^{-}|}\right) \left[
\sqrt{1 -\nu^2}
 \cosh \left(\frac{p_{4}\sqrt{1 -\nu^2}}
{2|\Delta_{0}^{-}|}\right) 
- \sinh \left(\frac{p_{4}\sqrt{1 -\nu^2}}
{2|\Delta_{0}^{-}|}\right)
\right], \quad \mbox{for} \quad 
0 \leq p_{4} \leq |\Delta_{0}^{-}| ,\label{sol-chi1}\\
\chi(p_{4}) &=& C_{2} 
\sin \left(\frac{\nu}{2} \ln\frac{\Lambda+|\Delta_{0}^{-}|}
{p_{4}}\right) , \quad \mbox{for} \quad 
|\Delta^{-}_{0}| \leq p_{4} \leq \Lambda , \label{sol-chi2}
\ea
\end{mathletters}
where $C_{i}$ ($i=1,2$) are the integration constants.
By matching the solutions at $p_{4} =|\Delta_{0}^{-}|$, we arrive
at the following condition that determines the value of the gap:
\be
\ln\frac{\Lambda+|\Delta^{-}_{0}|}{|\Delta^{-}_{0}|} =
\frac{2}{\nu}
\arctan\left(\frac{\sqrt{1-\nu^2}\coth(\sqrt{1-\nu^2}/2)-1}
{\nu}\right).
\label{eq-gap-con} 
\ee
This leads to the same (up to an overall constant of order one)
expression for the gap as in Eq.~(\ref{A-d}). Recall that the latter
was derived without taking the Meissner effect into account. We 
conclude, therefore, that the solution for the $\chi$-doublet NG 
boson is not sensitive to the screening 
due to the Meissner effect.

The analysis of the BS equation for the (pseudo-) NG bosons in the
$\lambda$-doublet and $\eta$-singlet channels is very similar to
the analysis for the $\chi$-doublet and we do not repeat it here.

\subsection{Massive diquarks}
\label{diff-eq-mass}

Let us consider the integral equation (\ref{BS-chi-mass}) for
the BS wave function of the massive $\chi$-doublet. It is
instructive to start with the analysis of this equation by
neglecting the Meissner effect at first. This is achieved by
substituting $c=0$. In this special case, the differential
equation reads
\be 
p_{4} \chi^{\prime\prime}(p_{4}) +\chi^{\prime}(p_{4})
+\frac{\nu^{2}}{4} K^{(\chi)}(p_{4}) \chi(p_{4}) =0 , \quad
\mbox{where} \quad  \nu=\sqrt{\frac{8\alpha_{s}}{9\pi}}, 
\label{chi-m1}
\ee
along with the same boundary conditions as in Eq.~(\ref{bound-c}).
The kernel [compare with Eq.~(\ref{K-chi})] is approximately
given by
\be 
K^{(\chi)}(p_{4}) =\left\{ \begin{array}{lll}
|\Delta_{0}^{-}|/\left(|\Delta_{0}^{-}|^{2}-M_{\chi}^{2}\right),
& \mbox{for} & 0 \leq p_{4} \leq z_{M}^{2}|\Delta_{0}^{-}|, \\
1/p_{4}, & \mbox{for} & 
z_{M}^{2}|\Delta_{0}^{-}| \leq p_{4} \leq \Lambda,
\end{array} \right.
\ee
where  $z_{M} =\sqrt{|\Delta^{-}_{0}|^{2}- M_{\chi}^{2}}
/|\Delta^{-}_{0}|$. 

The analytical solutions to the differential equation
in two qualitatively different regions are given by
\begin{mathletters}
\ba
\chi(p_{4}) &=& C_{1} J_{0} \left(
\frac{\nu}{z_{M}}\sqrt{\frac{p_{4}}{|\Delta^{-}_{0}|}}\right),
\quad \mbox{for} \quad 
0 \leq p_{4} \leq z_{M}^{2} |\Delta_{0}^{-}| ,\label{sol-m1}\\
\chi(p_{4}) &=& 
C_{2} \sin \left(\frac{\nu}{2} \ln\frac{\Lambda}{p_{4}}\right) ,
\quad \mbox{for} \quad  
z_{M}^{2} |\Delta^{-}_{0}| \leq p_{4} \leq \Lambda .
\label{sol-m2}
\ea 
\end{mathletters}
By matching these two solutions at $p_{4} = z_{M}^{2}
|\Delta_{0}^{-}| $, we obtain
\be 
\ln \frac{\Lambda}{z_{M}^{2}|\Delta^{-}_{0}|} = 
\frac{2}{\nu}\arctan\left(\frac{J_{0}(\nu)}{J_{1}(\nu)}\right)
+\frac{2\pi n}{\nu}, \quad n=1,2, \ldots .
\ee
By comparing this with the gap equation (\ref{eq-app-B}) (see also
Appendix~B in Ref.~\cite{us}), we derive the following spectrum of
massive diquarks in absence of the Meissner effect:
\be
M_{n} = |\Delta^{-}_{0}|
\sqrt{1-\exp\left(-\frac{2\pi n}{\nu}\right)},  
\quad n=1,2, \ldots .
\label{chi-spectrum}
\ee
Below we argue that none of these massive states survive after the 
Meissner effect is taken into account. In fact, this is almost
obvious when we notice that the BS wave functions that correspond
to the states with masses in Eq.~(\ref{chi-spectrum}) have rather
rich node structure in the far infrared region $0 < p \ll
|\Delta_{0}^{-}|$. Topologically, the $n$-th wave function has
exactly $n$ zeros. These $n$ zeros appear at 
\be
p_{4}^{(k)} = |\Delta_{0}^{-}| \exp\left[
-\frac{2}{\nu} \left(\pi k -\arctan\frac{J_{0}(\nu)}{J_{1}(\nu)}
\right) \right], \quad k=1,2,\ldots,n.
\ee
In the weakly coupled theory, $\nu\ll 1$, we find that $p_{4}^{(k)}
\ll |\Delta_{0}^{-}| $ for any $k$. This suggests that after taking
the Meissner effect back into consideration, the mentioned
structure of the nodes in the BS wave functions of the massive
$\chi$-diquarks would become impossible due to strong screening
effects of the gluons in the infrared region $0 < p_{4} \alt
|\Delta_{0}^{-}|$. 

To substantiate the claim of the previous paragraph, let us now
consider the equation where the Meissner effect is qualitatively
taken into account. We arrive at the following differential
equation: 
\be 
(p_{4}+|\Delta_{0}^{-}|) \chi^{\prime\prime}(p_{4})
+\chi^{\prime}(p_{4}) +\frac{\nu^{2}}{4} K^{(\chi)}(p_{4})
\chi(p_{4}) =0 , \quad \mbox{where} \quad 
\nu=\sqrt{\frac{8\alpha_{s}}{9\pi}}.
\label{dif-eq-chi}
\ee
The wave function should again satisfy the boundary conditions in
Eq.~(\ref{bound-c}). To get the estimate for the solution, we
consider the differential equation on the three adjacent intervals,
$0 \leq p_{4} \leq z_{M}^{2}|\Delta_{0}^{-}|$,
$z_{M}^{2}|\Delta_{0}^{-}| \leq p_{4} \leq |\Delta_{0}^{-}|$ and
$|\Delta_{0}^{-}| \leq p_{4} \leq \Lambda$, separately. The
corresponding analytical solutions read
\begin{mathletters}
\ba
\chi(p_{4}) &=& C_{0} 
\exp\left(\frac{p_{4}}{2|\Delta_{0}^{-}|}\right) \left[
\sqrt{z_{M}^{2} -\nu^2}
 \cosh \left(\frac{p_{4}\sqrt{z_{M}^{2} -\nu^2}}
{2z_{M}|\Delta_{0}^{-}|}\right) \right. \nonumber \\
&& \left. 
- z_{M} \sinh \left(\frac{p_{4}\sqrt{z_{M}^{2} -\nu^2}}
{2z_{M}|\Delta_{0}^{-}|}\right)
\right], \quad \mbox{for} \quad 
0 \leq p_{4} \leq z_{M}^{2} |\Delta_{0}^{-}| ,\label{sol-1}\\
\chi(p_{4}) &=& 
C_{1}\left[ G^{20}_{12}\left( \frac{p_{4}}{|\Delta_{0}^{-}|}
\left| \begin{array}{ll}1-\frac{\nu^2}{4} &  \\0 & 1
\end{array}\right. \right)\right. \nonumber \\
&& \left. 
+C_{2} \frac{p_{4}}{|\Delta_{0}^{-}|} \! ~_{1}\! F_{1}
\left(1+\frac{\nu^2}{4}, 2, -\frac{p_{4}}{|\Delta_{0}^{-}|} 
\right) \right],  \quad \mbox{for} \quad 
z_{M}^{2} |\Delta_{0}^{-}| \leq p_{4} \leq |\Delta_{0}^{-}|,
\label{sol-2}\\ 
\chi(p_{4}) &=& 
C_{3} \sin \left(\frac{\nu}{2} \ln\frac{\Lambda+|\Delta_{0}^{-}|}
{p_{4}}\right) , \quad \mbox{for} \quad 
|\Delta^{-}_{0}|\leq p_{4} \leq \Lambda , \label{sol-3}
\ea
\end{mathletters}
where $G^{20}_{12}$ is the Meijer's G-function. The solutions in the
first and the last regions are chosen so that they satisfy the
boundary conditions  in Eq.~(\ref{bound-c}).

We note that the ultraviolet asymptote (\ref{sol-3}) of the BS wave
function of a massive doublet coincides with that of the NG boson
(\ref{sol-chi2}). Moreover, this property is shared by all massive
states that exist, irrespective of the value of their mass. Now, 
unlike the wave functions of (pseudo-) NG bosons which have no nodes, 
the BS wave functions of massive excitations should have at
least one zero somewhere in the region $0 \leq p_{4} < \Lambda$.
Since, in agreement with our previous statement, there cannot be
any nodes in the ultraviolet region $|\Delta_{0}^{-}| < p_{4} <
\Lambda$, they might occur only
in the infrared, $0 \leq p_{4} \leq
|\Delta_{0}^{-}|$. 

By matching the logarithmic derivatives of the solutions
at $p_{4}=z_{M}^{2} |\Delta^{-}_{0}|$ and
$p_{4}=|\Delta^{-}_{0}|$, we obtain the two different
expressions for the integration constant $C_{2}$: 
\be
C_{2} = {\cal F}(z_{M}) \quad \mbox{and}  \quad
C_{2} = {\cal F}(1) \label{C2} ,
\ee
where the explicit form of function ${\cal F}(z)$ reads
\be
\frac
{\nu^2 z^2 \! ~_{1}\! F_{1}(1+\frac{\nu^2}{4}, 2, -z^2) 
+z\left[\sqrt{z^2 -\nu^2} 
\coth\left(\frac{z\sqrt{z^2 -\nu^2}}{2} \right)
-z\right] \left[ 2 \! ~_{1}\! F_{1}(1+\frac{\nu^2}{4}, 2, -z^2) 
-z^2(1 +\frac{\nu^2}{4}) 
\! ~_{1}\! F_{1}(2+\frac{\nu^2}{4}, 3, -z^2) \right]}
{2 z\left[\sqrt{z^2 -\nu^2} \coth\left(z\sqrt{z^2 -\nu^2}/2 \right)
-z\right] 
G^{20}_{12}\left( z^2 \left| \begin{array}{ll}-\nu^2/4 &  \\
                            0 & 0 \end{array}\right. \right)
-\nu^2 
G^{20}_{12}\left( z^2 \left| \begin{array}{ll}1-\nu^2/4 &  \\
                            0 & 1 \end{array}\right. \right)}.
\ee
Notice that we used the gap equation (\ref{eq-gap-con}) to derive 
the second expression in Eq.~(\ref{C2}). 

The spectrum of massive excitations (if any) should be determined by
the solutions of the equation ${\cal F}(z)={\cal F}(1)$ where $z<1$.
Note that the obvious solution $z=1$ corresponds to the no-node wave
function of the NG boson. By studying the equation numerically, we
find that there are no solutions which would correspond to wave
functions with nodes. However, in addition to $z=1$ solution, there
is another solution for $z<1$. This latter also corresponds to a
wave function {\em without} nodes in the whole region of momenta 
$0 \leq p_{4} < \Lambda$. In fact, its shape barely differs from
the wave function of the NG boson. In the spectral problem at hand,
however, one does not expect having two solutions with the same
no-node topology. Therefore, we believe that the extra solution is
an artifact of the approximations used. Its appearance apparently
results from two {\em different} splitting of the whole region of
momenta into separate intervals for $M=0$ and $M\neq 0$ cases. This
is also supported by the observation that, because of the Meissner
effect, the BS wave function in the doublet channel is always almost
a constant function (and, therefore, cannot have zeros) in the
infrared region $0 \leq p_{4} < |\Delta^{-}_{0}|$. This is seen
already from the integral version of the BS equation
(\ref{BS-chi-mass}). It is  natural that there is only one no-node
solution to the BS equation. Since the BS wave function of the NG
boson is such a solution, no other non-trivial solutions should
exist in the doublet channel.

The analysis of the BS equation for the $\lambda$-antidoublet is
similar and we do not repeat it here. The analysis for the singlet 
is presented in Subsec.~\ref{chi-m-doublet} in detail. There the
Meissner effect is qualitatively taken into account by considering
only the interaction that is mediated by the gluons of the unbroken
$SU(2)_{c}$ subgroup.

\end{document}